\documentclass[namedreferences]{solarphysics}

\usepackage[hyperref,optionalrh,showbiblabels]{spr-sola-addons} % For Solar Physics 
\usepackage{graphicx}        % For eps figures, newer & more powerfull
\usepackage{color}           % For color text: \color command
\usepackage{breakurl}        % For breaking URLs easily trough lines
\chardef\us=`\_

\begin{document}

\begin{article}

\begin{opening}

\title{Propagation of Solar Energetic Particles during Multiple Coronal Mass Ejection Events} 

\author[addressref={aff1},corref,email={silpoh@utu.fi}]{\inits{S.}\fnm{Silja}~\lnm{Pohjolainen}}%\sep
\author[addressref={aff2,aff3},email={fimubaa@utu.fi}]{\inits{F.}\fnm{Firas}~\lnm{Al-Hamadani}}%\sep
\author[addressref={aff2},email={eino.valtonen@utu.fi}]{\inits{E.}\fnm{Eino}~\lnm{Valtonen}}%\sep
%\author{\inits{}\fnm{}~\lnm{}\orcid{}}
%\author{P.~\surname{Author-a}$^{1}$\sep
%        E.~\surname{Author-b}$^{1}$\sep
%        M.~\surname{Author-c}$^{2}$      
%       }

%   \institute{$^{1}$ First affiliation
%                     email: \url{e.mail-a} email: \url{e.mail-b}\\ 
%              $^{2}$ Second affiliation
%                     email: \url{e.mail-c} \\
%             }
\address[id=aff1]{Tuorla Observatory, University of Turku, Piikki\"o, Finland}
\address[id=aff2]{Department of Physics and Astronomy, University of Turku, Turku, Finland}
\address[id=aff3]{Department of Physics, University of Basrah, Basrah, Iraq}

\runningauthor{Pohjolainen, Al-Hamadani, and Valtonen}
\runningtitle{Propagation of Solar Energetic Particles during CMEs}

%%%%%%%%%%%%%%%%%%%%%%%%%%%%%%%%%%%%%%%%%%%%%%%%%%%
%% Authors Names
%
%\author{Silja~\surname{Pohjolainen}$^{1}$\sep
%        Firas~\surname{Al-Hamadani}$^{2,3}$\sep
%        Eino~\surname{Valtonen}$^{2}$}
%% Affilations 
%
%\institute{$^{1}$ Tuorla Observatory, University of Turku, Piikki\"o, Finland\\ 
%$^{2}$ Department of Physics and Astronomy, University of Turku, Turku, Finland\\
%$^{3}$ Department of Physics, University of Basrah, Basrah, Iraq
%}
 
\begin{abstract}
We study solar energetic particle (SEP) events during multiple solar eruptions. 
The analysed sequences, on 24--26 November 2000, 9--13 April 2001, 
and 22--25 August 2005, consisted of halo-type coronal mass ejections (CMEs) 
that originated from the same active region and were associated with intense flares, 
EUV waves, and interplanetary (IP) radio type II and type III bursts. 
The first two solar events in each of these sequences showed SEP enhancements 
near Earth, but the third in the row did not. We observed that in these latter events 
the type III radio bursts were stopped at much higher frequencies than in the 
earlier events, indicating that the bursts did not reach the typical plasma 
density levels near Earth. To explain the missing third SEP event in each
sequence, we suggest that the earlier-launched CMEs and the CME-driven shocks 
either reduced the seed particle population and thus led to inefficient particle 
acceleration, or that the earlier-launched CMEs and shocks changed the propagation 
paths or prevented the propagation of both the electron beams and SEPs, so that 
they did not get detected near Earth even when the shock arrivals were recorded. 
\end{abstract}

\end{opening}

\section{Introduction}

Solar energetic particles (SEPs) are believed to be first accelerated in the 
lift-off phase of coronal mass ejections (CMEs). The SEP acceleration itself could 
be due to flaring processes, acceleration in coronal and/or interplanetary shocks, 
or it could be a combination of all \citep{trottet14}. Diffusive shock acceleration 
near the Sun of a population of seed particles with suprathermal energies has 
been suggested as one alternative \citep{tylka06}. Recently \citeauthor{kahler2014b} 
(\citeyear{kahler2014b}) studied if the background solar wind speed could affect the 
SEP event timescales, but no differences were found.  

Radio emission can be used as a probe to identify particle acceleration in the 
solar corona and interplanetary (IP) space. Solar type III radio bursts are 
observed when accelerated electrons travel outward in the solar corona and propagate 
along open magnetic field lines. They excite plasma oscillations at the local plasma 
frequency, which are then converted to radio waves at the fundamental and harmonic 
plasma frequencies \citep{dulk2000}. The fast frequency-drifting radio emission 
of the type III bursts is due to fast electron beams, typically originating from 
solar flares. 
Solar type II radio bursts show a much slower frequency drift and a wider emission 
band compared to the type III bursts. The emission mechanism for the type II 
bursts is basically the same as for the type IIIs, only that the accelerated electrons 
are believed to originate from propagating shocks. The emission features reflect 
the speed of the shock and the short propagation paths of shock-accelerated 
electrons, see review by, {\it e.g.}, \citeauthor{gopal2000} (\citeyear{gopal2000}). 
In the interplanetary space these shocks are usually CME driven.  

Dependences of SEP enhancements in multiple CME events have been investigated 
in several studies \citep{gopal2004, kahler2005, ding2013, kahler2014a}.
% Gopalswamy et al., (2004), Kahler and Vourlidas (2005), Ding et al. (2013), 
%and Kahler and Vourlidas (2014). 
\citeauthor{al-sawad09} (\citeyear{al-sawad09}) and  
\citeauthor{kocharov09} (\citeyear{kocharov09}) presented cases where two CMEs 
were launched close in time and both eruptions were associated with SEP enhancements. 
In both of these analyses it was suggested that the first shock weakens and becomes 
gradually transparent for the protons produced by the second eruption, behind the 
first CME.   

In this article we analyse triple events where the first two CMEs are associated with
SEP enhancements, but at the time of the third CME no SEP enhancement is observed.
We use solar radio data and {\it in-situ} magnetic field measurements to investigate 
how particle beams and shock fronts propagate during these time periods, to find if 
the particle propagation conditions or acceleration processes change in the course 
of the successive events.

\section{Data Analysis}
\label{sec:data}

The aim of this study is to investigate how energetic particles can propagate
during multiple solar eruptions. Our basic selection criterion for the analysed events
was that there had to be at least three consecutive CMEs that originated 
from the same active region within three--four days. 
Each of the CMEs had to be associated with an IP radio type II burst, since this type 
of burst can be considered as a sign of a propagating shock wave that is capable of 
accelerating particles. The {\it Wind}/{\it Plasma and Radio Waves} (WAVES) list of 
type II bursts and CMEs at 
\url{http://cdaw.gsfc.nasa.gov/CME\_list/radio/waves\_type2.html} 
was used to find event sequences that matched our requirements. The preliminary 
candidate list that consisted of 17 sequences of consecutive events is presented in 
Table \ref{table5} (Appendix). 

For the analysis, the CMEs had to be separable from each other so that their 
associated features could be investigated. In some cases the CMEs occurred so 
close in time to each other that their individual properties could not be separated 
(these could be called early CME interaction events). Energetic particles had to 
be observed with the first two CMEs, and this requirement removed from our list 
those events that were not well-connected to the field-lines near L1, 
where the {\it Energetic and Relativistic Nuclei and Electron} (ERNE: 
\citeauthor{torsti95}, \citeyear{torsti95}) instrument onboard the {\it Solar 
and Heliospheric Observatory} (SOHO) is located. Data gaps in the {\it in-situ} 
particle data also made the analysis impossible for some of the events. 
Eventually, three triple events were selected for closer inspection.   
 
The SOHO/{\it Large Angle and Spectrometric Coronagraph} (LASCO) CME catalog 
at \url{http://cdaw.gsfc.nasa.gov/} was used for finding the CME characteristics 
(width, speed, height-time evolution). The data for the associated IP radio type II 
events were taken from the {\it Wind}/WAVES \citep{bougeret95} catalog at 
\url{http://lep694.gsfc.nasa.gov/waves/data\_products.html},
prepared by Michael L. Kaiser. 
Flares and flare locations were checked from the NOAA National Geophysical Data 
Center (NGDC) flare listings at 
\url{http://www.ngdc.noaa.gov/stp/spaceweather.html}. 
X-ray data from the {\it Geostationary Operational Environmental Satellite} (GOES) 
were plotted using {\it Solarsoft}, in order to check the listings and to identify 
multiple flare events. 
The SOHO/{\it Extreme-Ultraviolet Imaging Telescope} (EIT: \citeauthor{delabou95}, 
\citeyear{delabou95}) images were searched to find EUV waves and dimmings.
 
The SOHO/ERNE proton data were then investigated  using the website at 
\url{http://www.srl. utu.fi/erne\_data/main\_english.html}, to find the 
associated SEP events. As supportive observations we used electron data from 
{\it Advanced Composition Explorer}/{\it Electron, Proton and Alpha Monitor}
(ACE/EPAM: \citeauthor{gold98}, \citeyear{gold98}) and SOHO/{\it Electron 
Proton Helium Instrument} (EPHIN: \citeauthor{muller95}, \citeyear{muller95}) 
from the SEPServer \citep{vainio13} at 
\url{http://server.sepserver.eu/index.php}. 
The ACE/{\it Magnetometer instrument} (MAG: \citeauthor{smith98}, \citeyear{smith98}) 
magnetic field measurements available at 
\url{http://www.srl.caltech.edu/ACE/ASC/level2/lvl2DATA_MAG.html} 
were used for assessing the local magnetic field conditions at the time of 
the SEP events.

For the verification of the shock arrival near Earth we checked the ACE, {\it Wind}, 
and SOHO/{\it Charge, Element, and Isotope Analysis System} (CELIAS) shock 
lists at \url{http://www.cfa.harvard.edu/shocks/} and at 
\url{http://umtof.umd.edu/pm/FIGS.HTML}. Interplanetary CME (ICME) arrival times 
were verified from \url{http://www.ssg.sr.unh.edu/mag/ace/ACElists/ICMEtable.html}.

To compare the CME heights with shock heights we calculated the radio source heights 
for the observed IP type II bursts. The observed radio frequency [$f$] in Hz depends on 
the local electron density [$n_e$] in cm$^{-3}$ by $f = 9000\sqrt{n_e}$. 
The methods of determining type II burst heights using atmospheric density 
models are described in, {\it e.g.}, \citeauthor{pohjolainen2007} 
(\citeyear{pohjolainen2007}). 
The hybrid electron density model of \citeauthor{vrsnak04} (\citeyear{vrsnak04}) was 
selected for the calculations, mainly because it merges the high-density low-corona 
models to the low-density IP models without breaks or discrepancies. 

\subsection{Events on 24--26 November 2000}

\begin{figure}[!t]
   \centering
   \includegraphics[width=0.95\textwidth]{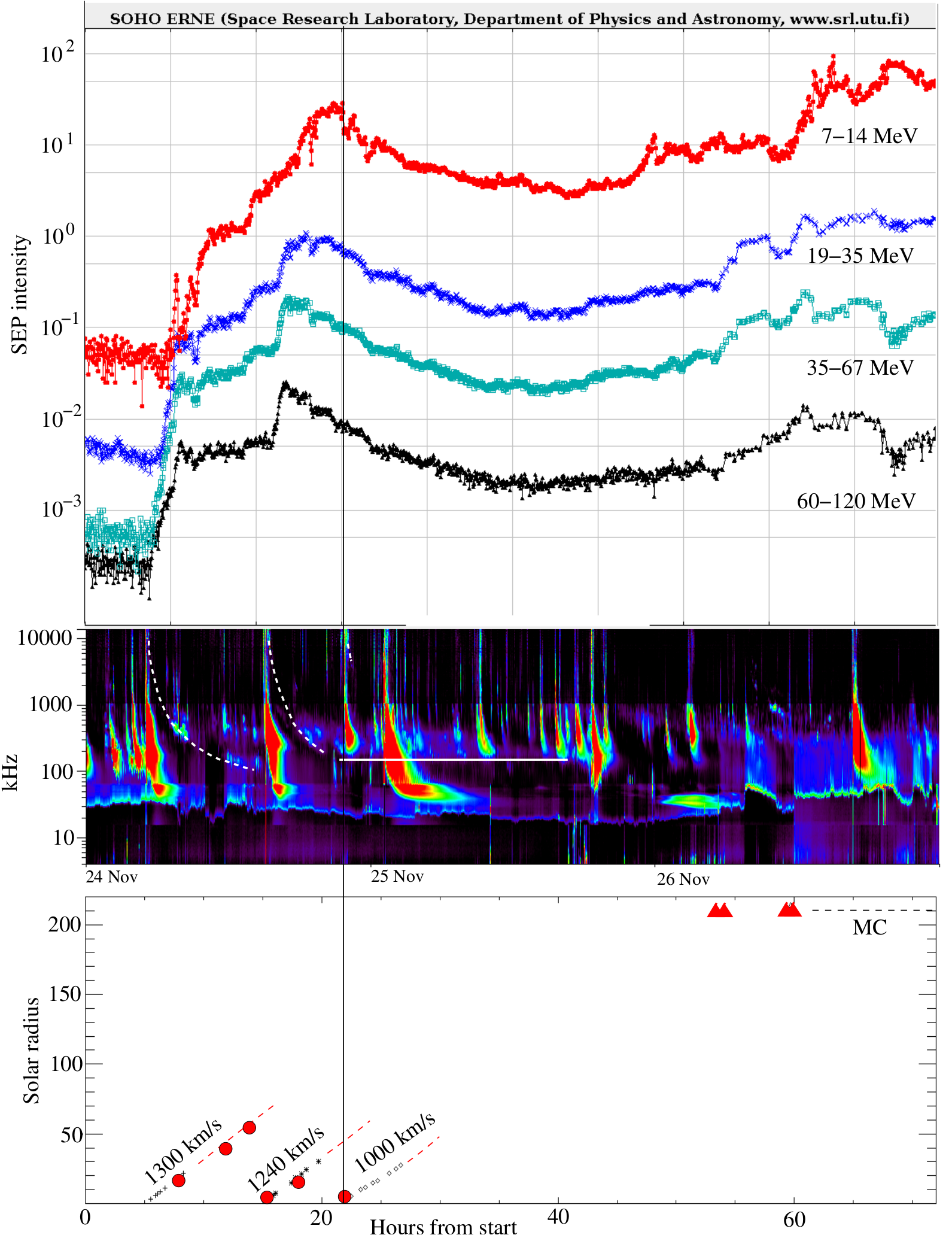}
   \caption{Solar events on 24--26 November 2000: Solar energetic 
proton intensities observed {\it in-situ} by SOHO/ERNE, in (cm$^2$\,s\,sr\,MeV)$^{-1}$ 
(top panel, energy range 7--120 MeV). 
The {\it Wind}/WAVES radio dynamic spectrum (middle panel, frequency range 14 MHz--4 kHz)
shows the associated IP type II bursts (the lanes are enhanced with 
white dashed lines) and the period of type III burst cut-offs (white horizontal line).
The type III burst in the early hours of 25 November that shows no cut-off    
originated from an active region located near the East limb.
The height-time plot (bottom panel) shows the associated CMEs 
and their estimated initial velocities, the calculated type II heights 
(red circles), and the times of shock arrival near L1 and Earth (red triangles).
The black vertical line marks the time after which a SEP enhancement would
have been expected to appear. The black dashed line marks the MC-like structure
reported by Wang, Wang, and Ye (2002). 
}  
   \label{24nov2000}%
   \end{figure}

On 24 November 2000 three GOES X-class flares were observed originating from
the same active region, NOAA 9236. The flares were associated with fast 
halo-type CMEs and their approximate initial speeds are listed in 
Table \ref{table:1}. 
Two C-class flares (C4.2 and C2.4) were reported between the X2.0 and
X2.3 flares from the same active region. Between the X2.3 and
X1.8 events, no flares were reported from this active region.
All three flare-CME events were associated with an IP radio type II burst, as
seen in the {\it Wind}/WAVES dynamic spectrum in Figure \ref{24nov2000}. The type II 
emission lanes are enhanced with white dashed lines. 

\begin{table}[!t]
\caption{Solar events observed during 24--26 November 2000.}             
\label{table:1}              
\begin{tabular}{l c c c c l l}
\hline       
Date         & Flare & Flare  & Flare  & CME    & Shock                & Shock \\ 
             & start & site   & class  & speed  & arrival and          & arrival and \\ 
             &       &        &        &        & speed*               & speed*\\
             &       &        & GOES   &        & ACE                  & {\it Wind}\\
             & UT    &        &        & km\,s$^{-1}$   & km\,s$^{-1}$   & km\,s$^{-1}$\\            
\hline
2000 Nov. 24 & 04:55 & N20W06 & X2.0  & 1300    & Nov 26, 05:00\,/462  & 05:32\,/520\\
2000 Nov. 24 & 14:51 & N22W08 & X2.3  & 1240    & Nov 26, 11:24\,/632  & 11:43\,/666\\
2000 Nov. 24 & 21:43 & N21W14 & X1.8  & 1000$^{1}$ &                      &           \\ 
\hline                  
\end{tabular}\\
\mbox{*Shock} velocity is given after the slash mark.\\
$^{1}$Estimated speed using the last two CME height observations is 1240 km/s.\\  
{\it Wind} near Earth, distance to ACE $\sim$ 950\,000 km.\\
\end{table}

The first IP type II burst was rather complex in its beginning, but 
it was very intense, with a clear narrow-band emission lane at frequencies 
below 1 MHz. The calculated type II source heights fit well with the CME heights,  
as shown by the height-time plot in Figure \ref{24nov2000} (bottom panel, 
type II heights are indicated with red filled circles). 
The second type II burst was comprised of narrow-band short-duration bursts, with 
a separate, more diffuse emission lane at lower frequencies. We used this lower 
frequency lane for the height calculations, and the obtained heights are again 
in good agreement with the corresponding CME heights. 
The third type II burst was visible only down to 3 MHz, and the calculated 
type II heights match with those of the CME in its early phase. 

All three flare-CME events were also associated with radio type III bursts.
The first two continued down to the plasma density level near Earth, to about 30 kHz,
but the third one ended at about 200 kHz. This plasma frequency corresponds to 
an electron density of 494 cm$^{-3}$, which in standard atmospheric density 
would be at a distance of about 30 R$_{\odot}$ from the Sun center. 
At the time of the type III cut-off start, the leading front of the 
first CME was estimated to have a height of about 100 R$_{\odot}$ and the second 
CME front a height of about 40 R$_{\odot}$. The radio dynamic spectrum shows little 
difference between the frequency of the second type II burst and the type III 
end-frequency near the time of the cut-off start, suggesting that the type III 
cut-off happens approximately at the height where the shock front of the second CME 
is located.

The strong type III burst after the third event, in the early hours of 25 November,   
originated from a flare at N07E50, with an eastward-directed CME. Hence it 
had a completely different magnetic connectivity to Earth compared to the 
three earlier events. During the most part of 25 November, the type III bursts 
associated with flares in AR NOAA 9236 had similar shorter propagation paths 
and their emission ended near 200 kHz.

Before the first X-class flare at 04:55 UT on 24 November, several radio 
type III bursts occurred originating from the same active region NOAA 9236, and 
were observed to have a cut-off at the second harmonic of the local plasma
frequency (2f$_{p}$). Two local plasma frequency emission lanes 
($\sim$20 and $\sim$40 kHz) were clearly visible in the dynamic spectra during 
several days before the type III burst activity on 23 November. However, the 
strong type III burst emission associated with the first X-class flare at 04:55 UT 
was able to reach the fundamental plasma frequency level and by that time also 
the second harmonic lane had disappeared from the spectrum.  

The ERNE {\it in-situ} observations of energetic protons show two enhancements,
the first one associated with the start of the first CME-type II event and the second 
with the start of the second CME-type II event. At the time of the start of 
the third CME-type II event the SEP flux was decreasing at all energies 
(Figure \ref{24nov2000}, top panel).  

At least two shocks were observed to arrive near Earth, recorded by
ACE (near L1) and {\it Wind} (near Earth) on 26 November. They are listed in 
Table \ref{table:1} with the shock arrival speeds. In Figure \ref{24nov2000} 
the  shock arrival times are indicated with red triangles in the height-time
plot in the bottom panel. Also the SOHO/CELIAS list contains two possible shocks 
on 26 November. The arrival time of the first one (07:15 UT) is two hours later 
than the ACE and {\it Wind} shocks but the second one (11:26 UT) fits well with
the other observations. The travel times and CME speeds suggest that these 
shocks could be associated with the first two CMEs and, with some caution, 
also with the third CME. For example \citeauthor{wang2002} (\citeyear{wang2002})
have suggested that the second shock could have been produced by several shocks
merging. 

\citeauthor{wang2002} (\citeyear{wang2002}) reported an IP magnetic cloud (MC) -like 
structure observed by the {\it Wind} satellite starting at about 13:00 UT on 26 November. 
They propose it to be the result of at least four halo-CMEs, of which the first 
three are the homologous CMEs analysed here. Later on, there are two listed ICME 
passages on the 27 and 28 November. \citeauthor{cane2003} (\citeyear{cane2003}) have 
marked them as multiple CME events, the 27 November ICME as the result of the 
CMEs on 24 November and the 28 November ICME as the result of several CMEs that 
occurred during 25--26 November.

%__________________________________________________________________

%%%%%%%%%%%%%%%%%%%%%%%%%%%

\subsection{Events on 9--13 April 2001}

\begin{figure}[!t]
   \centering
   \includegraphics[width=0.95\textwidth]{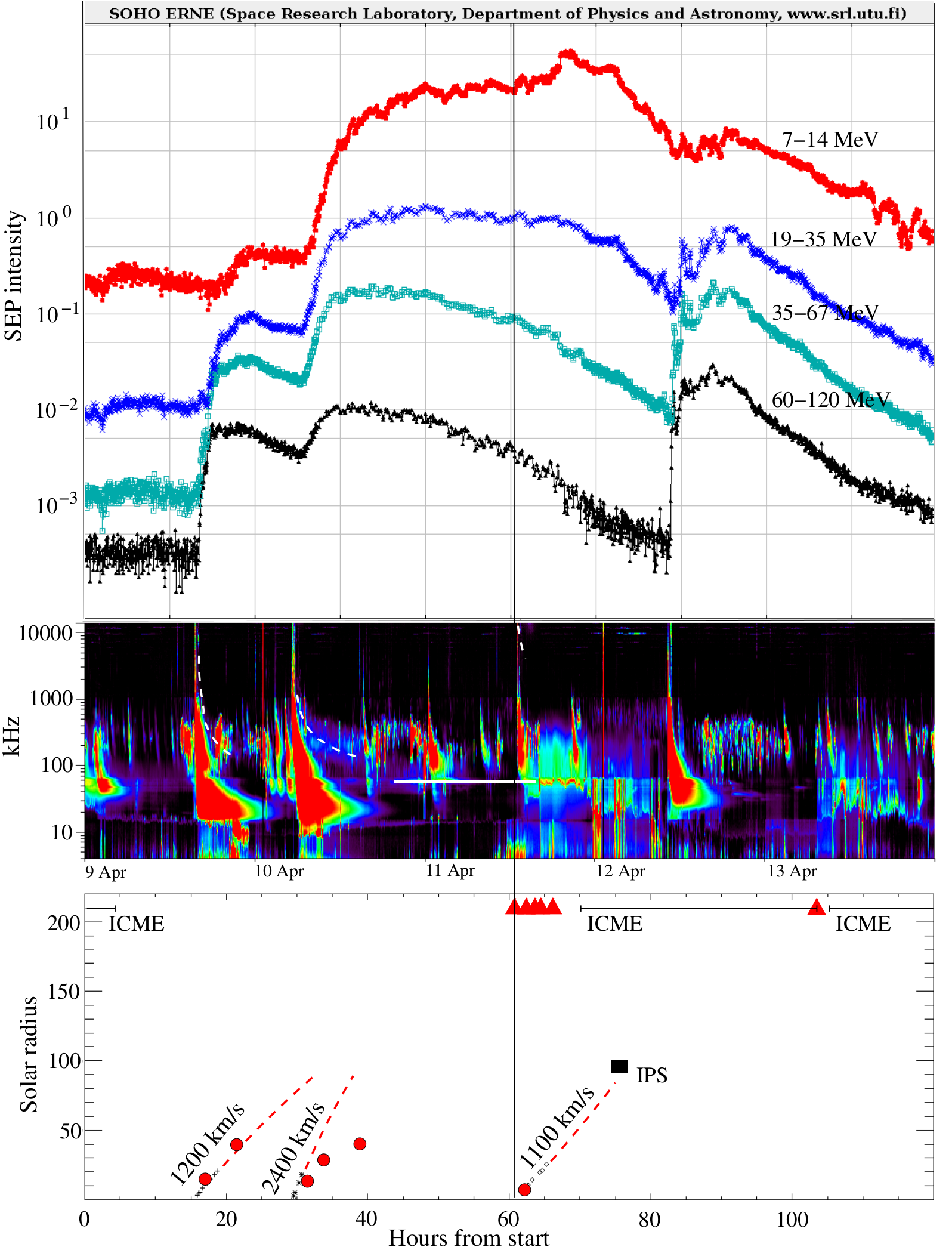}
   \caption{Solar events on 9-13 April 2001, panels, symbols, and units  
    are the same as in Figure \ref{24nov2000}. The black box marks the estimated
    CME height from IPS measurements reported by Iju et al. (2013).
    }  
   \label{9apr2001}%
   \end{figure}

In April 2001 three consecutive CME and IP type II burst-associated flares 
originated from the same active region NOAA 9415, observed on 9, 10, and 11
April. The flares were GOES class M7.9, X2.3, and M2.3, respectively. The flares 
and the high-speed halo-type CMEs are listed in Table \ref{table:2}. 
Between the M7.9 and X2.3 flares, one C2.1 flare was reported from active 
region 9415. Between the X2.3 and M2.3 flares there was one C2.7 flare.
The associated IP type II bursts are visible in the {\it Wind} WAVES dynamic spectrum 
in Figure \ref{9apr2001} (enhanced with white dashed lines). 

\begin{table}[!t]
\caption{Solar events observed during 9--13 April 2001}             
\label{table:2}               
\begin{tabular}{l c c c c l l}
\hline       
Date         & Flare & Flare  & Flare  & CME    & Shock               & Shock \\
             & start & site   & class  & speed  & arrival and         & arrival and\\ 
             &       &        &        &        & speed*              & speed*\\
             &       &        & GOES   &        & ACE                 & {\it Wind}       \\
             & UT    &        &        & km\,s$^{-1}$  & km\,s$^{-1}$   & km\,s$^{-1}$\\            
\hline
2001 Apr. 9   & 15:20 & S21W04 & M7.9  & 1200    & Apr 11, 13:14\,/613 & 14:09\,/673\\  
2001 Apr. 10  & 05:06 & S23W09 & X2.3  & 2400    & Apr 11, 15:28\,/731 & 16:18\,/828\\
             &       &        &       &         &                        & 18:18\,/830\\
2001 Apr. 11  & 12:56 & S22W27 & M2.3  & 1100    & Apr 13, 07:06\,/914 &           \\ 
\hline                  
\end{tabular}\\
\mbox{*Shock} velocity is given after the slash mark.\\
{\it Wind} near Earth, distance to ACE $\sim$ 1\,400\,000 km. \\
\end{table}

The radio dynamic spectrum shows a short-duration IP type II burst 
at 12--7 MHz on 9 April but there was also intermittent emission observed 
at lower frequencies. We measured the type II burst heights using this lower, 
less intense emission lane. 
The type II source heights are in good agreement with the heights of the 
associated CME (Figure \ref{9apr2001}, bottom panel). The next IP type II 
burst on 10 April had a very wide and diffuse emission band, and the height 
calculations were done using the emission lane center. Similar to some of the 
wide-band events reported in  \citeauthor{pohjolainen13} (\citeyear{pohjolainen13}), 
the velocity of the type II burst source decreased rapidly and the burst showed 
a different height-time evolution compared to the associated CME. The third IP 
type II burst on 11 April had a narrow band and it was visible for less than two 
hours from the beginning of the flare-CME event. The type II heights matched those 
of the CME in its initial phase.  
  
All three flare-CME events were also associated with radio type III bursts.
The first two type IIIs continued down to the plasma level near Earth, to 
about 30 kHz, but the third ended near 60 kHz. In standard atmospheric density 
this would mean a distance of about 90 R$_{\odot}$ from the Sun center. 
At the time of the cut-off, the leading edges of the earlier two CMEs were 
both estimated to be at a height of about 150 R$_{\odot}$. 
The third type III event was also associated with prolonged enhanced radio
emission observed around 50--100 kHz. This period coincides with the detection
of several shock arrivals near Earth (indicated by red triangles in 
Figure  \ref{9apr2001}), so it could be associated with local shocks caused by the
earlier two CMEs that had merged during their propagation.  

For the third event there are interplanetary scintillation (IPS) measurements
of the CME height \citep{iju13}.  The CME was estimated to be at a height of 
95 R$_{\odot}$ at 03:34 UT on 12 April. This position is in line with the CME/type II 
heights from the start to the arrival of the shock near Earth on 13 April 07:06 UT
(Figure \ref{9apr2001}, bottom panel). 

The {\it in-situ} observations of energetic particles show two enhancements, the 
first one associated with the first CME-type II start and the second with the 
second CME-type II start. At the time of the third CME-type II start the 
SEP flux curve was decreasing at higher energies. We note that the next 
flare-CME event from active region NOAA 9415 was on 12 April and was 
again associated with a SEP enhancement, but this time it was more pronounced 
in the higher energies. This flare-CME event was associated with a 
metric type II burst but no IP type II emission was observed. 
Between the 12 April flare and the preceding, analysed M2.3 flare on 
11 April three other flares (C7.7, B7.2, and M1.3) occurred in the same 
active region.
As the 12 April SEP event was observed inside the earlier-launched ICMEs,  
there could have been a direct connection to the active region.

Several shocks were observed to arrive near Earth, recorded by ACE (near L1) 
and {\it Wind} (near Earth) on 11 and 13 April (Table \ref{table:2} and Figure \ref{9apr2001}, 
bottom panel). SOHO/CELIAS also observed the first two shock arrivals on 11
April (13:02 and 15:18 UT), as well the shock arrival on 13 April (07:05 UT).
The shocks on 11 April seem to be associated with the first two CME--type II events, 
which probably had merged together during propagation. The shock on 13 April could be 
connected with the third CME-type II event, as the IPS observation of the CME is 
located in between the start of the CME and the shock arrival. However, we cannot 
rule out the possibility that the 13 April shock was associated with the 
decelerating, wide-band type II burst observed on 10 April. 

Several ICME passages have been reported for these dates \citep{cane2003}. The
updated list of Near-Earth Interplanetary Coronal Mass Ejections, compiled
by Ian Richardson and Hilary Cane, online at 
\url{http://www.srl.caltech.edu/ACE/ASC/DATA/level3/icmetable2.htm},  
reports an earlier ICME passage that started on 8 April and continued 
until 04:00 UT on 9 April. The next passage started at 22:00 UT on 11 April  
and ended at 07:00 UT on 13 April. The following ICME passage started two hours
later on the same day and ended on 14 April.    

\begin{table}[!t]
\caption{Solar events observed during 22--25 August 2005}             
\label{table:3}              
\begin{tabular}{l c c c c l l}
\hline     
Date         & Flare & Flare  & Flare    & CME     & Shock                & Shock\\
             & start & site   & class    & speed   & arrival and             & arrival and\\ 
             &       &        &          &         & speed*           & speed*\\
             &       &        & GOES     &         & ACE                  & {\it Wind} \\
             & UT    &        &          & km\,s$^{-1}$ & km\,s$^{-1}$      & km\,s$^{-1}$\\            
\hline
2005 Aug. 22  & 00:44 & S11W54    & M2.6  & 1190    &                      &        \\ 
2005 Aug. 22  & 16:46 & S13W65    & M5.6  & 2400    & Aug 24, 05:45\,/581 & 05:35\,/591\\
             &       &            &       &         & Aug 24, 08:20\,/795 & 08:24\,/525\\ 
2005 Aug. 23  & 14:19 & S14W75$^{1}$ & M2.7  & 1900  & Aug 25, 13:00$^{2}$   &       \\
\hline                  
\end{tabular}\\
\mbox{*Shock} velocity is given after the slash mark.\\
$^{1}$Estimated flare location on the disk S14W75--80, flare loop top above the limb.\\
$^{2}$See Papaioannou et al. (2009).\\
{\it Wind} near ACE at L1.\\
\end{table}

\subsection{Events on 22-25 August 2005}
\label{sec:22aug}

\begin{figure}[!ht]
   \centering
   \includegraphics[width=0.95\textwidth]{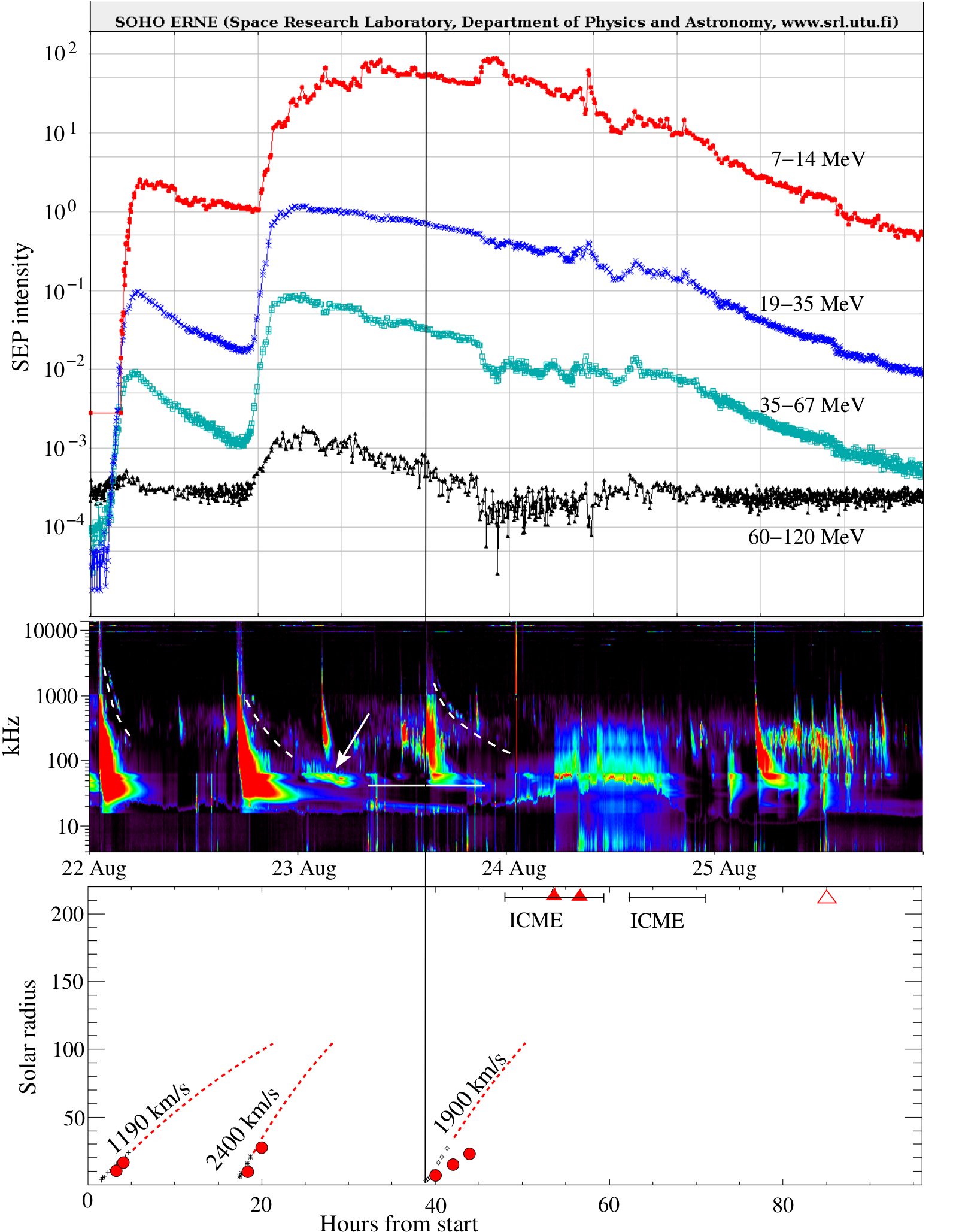}
   \caption{Solar events on 22--25 August 2005, panels, symbols, and units 
    are the same as in Figure \ref{24nov2000}. The white arrow in the radio 
    dynamic spectrum points to the blob-like structures discussed in the text. 
    The red unfilled triangle indicates the time of a disturbance reported by
    Papaioannou et al. (2009). }  
   \label{22aug2005}%
   \end{figure}

Three consecutive GOES M-class flares were observed originating from
the same active region NOAA 10798 on 22--23 August 2005. 
The flares were associated with fast halo-type CMEs and their approximate initial 
speeds are listed in Table~\ref{table:3}. 
Between the M2.6 and M5.6 flares two B-class flares (B5.2 and B5.3)
were reported from the active region. Between the second
and third flare in Table 3, no flares were reported from this active
region.
All three flare-CME events were also associated with an IP type II burst, see the 
{\it Wind} WAVES dynamic spectrum in Figure \ref{22aug2005}. 

The first IP type II burst on 22 August had a strong, narrow emission lane which 
could be followed from 2 MHz down to 400 kHz. The calculated type II burst source heights 
are in good agreement with the CME heights, see Figure \ref{22aug2005}, bottom panel. 
The next type II burst on the same day was more chaotic, with patches of emission
superposed on weaker emission. We used the diffuse emission lane center for the
height calculations and the match with the CME heights is again good. We note that
from 01:00 to 06:00 UT on 23 August there are enhanced radio emission features at
80--50 kHz (they appear as blob-like structures in the radio dynamic spectrum,
indicated with a white arrow in Figure \ref{22aug2005}) which could be signatures 
of CME interaction. The corresponding distance from the Sun 
center is about 65--100 R$_{\odot}$, which is a realistic height when looking at the 
height-time data of the first two CMEs and the related type II shocks.    
The third type II burst on 23 August showed very clear and narrow emission bands  
both at the fundamental and harmonic plasma frequencies. The lanes continued down 
to about 300 kHz before they got mixed with other emission features. The calculated 
type II heights, however, are considerably lower than the estimated CME heights 
after the event start, similar to the wide-band type II shock on 10 April 2001 which 
was described earlier.

The three flare-CME events were also associated with radio type III bursts.
The first two type IIIs continued down to the plasma density level near Earth, 
to about 30 kHz, but the third ended near 50--80 kHz (corresponding to the same 
distance as for the earlier possible CME interaction signatures). The end frequency 
of the third type III burst is not easily determined as there is also additional 
enhanced emission present (non-drifting and slowly-drifting). 
The estimates of the leading front heights of the earlier two CMEs at the time
of the type III cut-off start give heights of about 180--160 R$_{\odot}$, respectively, 
but as the first CME was decelerating and the second CME was accelerating they could 
have already merged, as suggested by the blob-like emission structures at 80--50 kHz 
observed some hours earlier.

The {\it in-situ} observations of energetic particles show two enhancements,
first one associated with the first CME-type II start and the second with the 
second CME-type II start (Figure \ref{22aug2005}, top panel). At the time of the 
third CME-type II start, the SEP flux was going down and decreasing in all energy 
ranges.

The first two close-in-time shock arrivals were recorded by ACE and {\it Wind} 
(both near L1) on 24 August (Table \ref{table:3} and Figure \ref{22aug2005}, 
bottom panel). They seem to be associated with the first two CME-type II events
that were launched on 22 August. SOHO/CELIAS lists only one shock (05:34 UT)
which is near the times of the first shock reported by ACE and {\it Wind}. For the 
third CME --type II event that was launched on 23 August \citeauthor{papa09} 
(\citeyear{papa09}) reported a disturbance observed by ACE on 25 August at 13:00 UT. 
A minor geomagnetic storm was observed to follow at 15:00--18:00 UT. 

The Richardson and Cane list of Near-Earth Interplanetary Coronal Mass Ejections 
contains two ICME passages for 24 August. The first started at 00:00 UT and 
ended at 11:00 UT, the second started at 14:00 UT and ended at 23:00 UT. 
A geomagnetic storm was also reported at 11:00 UT \citep{uwamahoro2012}. 

\subsection{SEP Characteristics and Interplanetary Conditions}

The results from the SEP data analysis are summarized in Table \ref{table:4}.
The table also gives the flare start times and the CME heights at the estimated 
times of first SEP injection. 

\subsubsection{24 November 2000}

The ERNE {\it in-situ} observations of solar protons at energies 1--130~MeV showed 
two enhancements on 24 November 2000 (Figure \ref{24nov2000}). 
These events occurred on the tail of the very large and long-duration event 
on 8 November. At the time of the first event onset, the background was
approximately at the quiet-time level only at the highest energies ($>$50 MeV). 
The onset of the event was, however, clearly visible at all energies up to 100 MeV. 
A velocity dispersion analysis (VDA) indicated a release time of protons at the Sun 
at 05:21 UT $\pm$ 14 minutes with an apparent path length of 2.09 $\pm$ 0.19 AU 
\citep{vainio13}. Thus, the release time of the particles is well associated with 
the first flare (04:55~UT) and the CME, which at the time of the estimated 
particle release time was at the height of 2--3 R$_{\odot}$. 

The onset of the second event was strongly masked by the first event, but was well
visible at the highest energies to beyond 100 MeV. At energies below $\sim$\,20 MeV 
it was hardly distinguishable from the still rising intensities related to the 
first event roughly 10 hours before. The VDA of this event, by using 
only the highest energy channels (17--130 MeV), give a release time at 15:27 UT 
with an apparent path length of 1.5 AU. The release time agrees reasonably well with 
the flare timing (14:51 UT), with a CME height of about 3 R$_{\odot}$ at the time 
of the particle release. 

At the time when the third flare was observed (21:43 UT) and when the associated CME 
was at a height of about 3 R$_{\odot}$ ($\sim$\,22 UT), no sign of a proton enhancement 
was observed at any energy channel. Later on,  on 25 November at about 15:00 UT, the 
intensities started rising slowly, probably related to the fast eastern halo CME at 
01:32 UT on 25 November. On 26 November small SEP enhancements were visible at lower 
energies, which may be related to the shock arrivals indicated in Figure \ref{24nov2000}.

\begin{table}[!t]
\caption{SEP characteristics and other observed features}             
\label{table:4}               
\begin{tabular}{l c l c r l l}
\hline       
Date         & Flare & SEP               & Time from   & CME         & Path   & EUV \\
             & start & injection         & flare start & height      & length & wave \\
             & (UT)  & (UT)              & (hours)     & (R$_{\odot}$)  & (AU)  & (UT)  \\
\hline
2000 Nov. 24  & 04:55 & 05:21\,$\pm$\,14 & 0.43  & $\sim$2.5  & 2.09\,$\pm$\,0.19    & 05:12\\
2000 Nov. 24  & 14:51 & 15:27            & 0.60  & $\sim$3.0  & 1.5 (highest energy) & 15:12\\
2000 Nov. 24  & 21:43 & --               &       &            &                      & 22:00 \\ 
\hline
2001 Apr. 9   & 15:20 & 16:01            & 0.68  & $\sim$4.0  & 1.2 (fixed length)   & 15:36 \\
2001 Apr. 10  & 05:06 & 06:48\,$\pm$\,11 & 1.70  & $\sim$19.0 & 1.38\,$\pm$\,0.15    & 05:24\\
2001 Apr. 11  & 12:56 & --               &       &            &                      & 13:13\\ 
\hline
2005 Aug. 22  & 00:44 & 01:29\,$\pm$\,9  & 0.75  & $\sim$4.0  & 1.62\,$\pm$\,0.12    & 01:13\\ 
2005 Aug. 22  & 16:46 & 17:30            & 0.70  & $\sim$6.0  & 1.2 (fixed length)   & 17:12\\
              &       & 18:22\,$\pm$\,9  & 1.60  & $\sim$16.0 & 2.06\,$\pm$\,0.09    & \\
2005 Aug. 23  & 14:19 & --               &       &            &                      & 14:48\\
\hline                  
\end{tabular}
\end{table}

We also inspected the electron data from ACE/EPAM and SOHO/EP\-HIN. 
We used ACE/EPAM deflected electron channels at 0.103--0.175~MeV and 0.175--0.315~MeV 
and SOHO/EPHIN electron channels at 0.25--0.7~MeV and 0.67--3 MeV. 
Electrons associated with the two first flare-CME events were clearly visible, 
but no enhancement was observed at the time of the third event. We also used 
the 0.175--0.315 MeV and 0.67--3 MeV channels of ACE/EPAM and SOHO/EPHIN,
respectively, to determine the release times of electrons by assuming
a nominal path length of 1.2 AU for the particles. The release times
obtained were 05:32 UT (EPAM) and 05:12 UT (EPHIN) for the first event
and 15:32 UT (EPAM) and 15:33 UT (EPHIN) for the second event. These
are consistent with the proton release times (Table \ref{table:4}).

In order to assess effects of interplanetary conditions on propagation 
of particles possibly accelerated during the third flare-CME event, we 
investigated ACE/MAG magnetic field data. We studied both the large-scale 
sector structure of the interplanetary magnetic field and possible smaller scale 
flux-rope structures at the time of the three events which could change the 
propagation paths of protons from the third event so that they would remain 
undetected at Earth. We used measurements of magnetic field components in 
Geocentric Solar Ecliptic (GSE) coordinates, with time resolution ranging 
from 1 hour to 16 seconds, as appropriate for the studied structures. 
A polarity change of the interplanetary magnetic field (IMF), from
away to towards the Sun, occurred on 23 November just before the
first event of the November 2000 sequence, and the polarity reversed 
again on 1 December. 
During the early hours of 24 November, there were periods of strong
fluctuations in the magnetic field direction and a slow rotation of
the field between about 07:15 UT and 09:30 UT, possibly related to a
flux-rope structure. At the time of all three flare-CME and SEP events
of 24 November, the magnetic field had the same polarity with
approximately the same direction close to the nominal Parker spiral. 
However, these {\it in-situ} measurements do not reveal the configuration 
of the IMF between the Sun and Earth at the time of particle
propagation through the IP space. Obviously, the shock structure
related to the first CME did not prevent protons of the second event
reaching Earth. However, the first and second shocks together might
have blocked the way of the protons associated with the third
flare-CME event. These shock structures and a magnetic cloud were
detected at Earth on 26 November (see Figure \ref{24nov2000}).

\subsubsection{9--11 April 2001}

The {\it in-situ} observations of energetic particles showed two enhancements on
9--10 April 2001 (Figure \ref{9apr2001}). These enhancements were preceded by a 
series of three major SEP events starting at the end of March, and the
background was high at all energies $<$50 MeV. Due to the background, the first 
enhancement was not observable below 5 MeV, but at higher energies it was visible 
up to $\sim$\,90~MeV. The initial VDA did not give a reasonable result \citep{vainio13}. 
In this event, the VDA gives an unreasonably long path length (4--7 AU, depending 
on which energy channels are used) indicating that the assumptions behind the VDA 
are not valid. By inspecting the intensity-time profiles, the earliest onset 
time of all energy channels was found to be at 16:18 UT at 90.5 MeV. By assuming 
a path length of 1.2 AU, this would result in a release time at 15:53 UT at the Sun 
(plus 8 minutes to be comparable with electromagnetic observations). 
This proton release time was consistent with the derived
electron release times of ACE/EPAM (16:02 UT) and SOHO/EPHIN (16:00 UT).
This is also in reasonable agreement with the flare start time (15:20 UT) and 
the CME observation at the height of $\sim$\,4 R$_{\odot}$. 

During the second SEP enhancement, the intensities below 10 MeV rose two orders of 
magnitude above the level of the first one, and at $\sim$\,35 MeV they were still 
an order of magnitude above the previous one. At 90 MeV this second event was not
anymore very pronounced, but still rose to the peak level 
($\sim$\,5$\times$10$^{-3}$ cm$^{-2}$ s$^{-1}$ sr$^{-1}$ MeV$^{-1}$) 
of the first already decaying event at this energy.
For this second event, the VDA gives a release time at 06:48 UT $\pm$ 11 minutes 
and an apparent path length of 1.38 $\pm$ 0.15 AU. 
Due to the high background and slowly rising profiles it is, however, difficult 
to observe the first-arriving particles, thus increasing the uncertainty of the VDA results. 
The obtained SEP release time is quite late compared to the flare time (05:06 UT)
and at that time the second CME was already at the height of $\sim$\,19 R$_{\odot}$. 
The obtained electron release times associated with the second flare-CME event 
were 05:58 UT (EPAM) and 06:26 UT (EPHIN), which are somewhat earlier than the 
derived proton release time.
No other major flares or fast CMEs were observed during the early hours of 10 April. 
Therefore, although the VDA results would be unreliable, it seems reasonable to 
assume that the cause of the observed particle enhancement at 06:40-06:50 UT is 
the flare-CME event of 10 April at 05:06/05:30 UT.

As in November 2000, no particle enhancement was observed which could have been 
associated with the third flare-CME-type II burst events. 
On 11 April from 13:00 to 14:00 UT, after the flare and while the CME was still 
below the height of about 6 R$_{\odot}$, the proton intensities above $\sim$\,25 MeV 
were slowly decaying. At the lowest energies the intensities were slowly rising, 
probably due to the approaching shocks (Table \ref{table:2}), peaking at 
$\sim$\,19:20 UT after the last reported shock in the {\it Wind} observations. 
A small drop in intensities was also observed at middle energies, coincident with 
the arrival of the ICME (Figure \ref{9apr2001}).
A slow enhancement in ACE/EPAM and SOHO/EPHIN electron intensities,
most probably associated with the shock arrival, was also observed
starting at about 13:00 UT.

Again we studied the magnetic field data to find a possible reason for not 
detecting protons associated with the third flare-CME event, due to interplanetary 
conditions. During all three events SOHO was located in the same polarity sector 
with a field direction close to the nominal Parker spiral. No flux-rope structures were 
observed. The magnetic field strength remained relatively stable during the events, 
although right after the third flare-CME event there was a sudden increase 
particularly in the x-component of the magnetic field, related to the shock arriving   
at SOHO. This shock structure might have changed the path of particles, possibly 
associated with the third flare-CME event, diverting them away from SOHO.

\subsubsection{22--23 August 2005}

Two proton enhancements were observed by ERNE on 22 August 2005. The first
one occurred at quiet-time background. The intensities rose rapidly by
approximately two orders of magnitude, up to 50 MeV energies (Figure \ref{22aug2005}). 
At 72 MeV this event was not anymore well visible. For this event, the VDA 
gives a release time at 01:29 UT $\pm$ 9 minutes, with an apparent path length of
1.62 $\pm$ 0.12 AU. 
The derived electron release times are slightly earlier: 01:09 UT
(0.175--0.315 MeV, EPAM) and 01:04 UT (0.67--3 MeV, EPHIN). The proton
release time, however, matches reasonably well with the flare start 
time (00:44 UT). 
The associated CME was then at the height of $\sim$\,4 R$_{\odot}$. 

The separation of the two particle events was only about 15 hours, so the second 
event occurred on the tail of the first one. Excluding the lowest energies 
($<$3 MeV), this event was, however, well distinguishable with the intensities 
exceeding the peak intensities of the first event by at least one order of magnitude 
at all energies up to 72~MeV, and was still visible at 90 MeV. The release
time was estimated to be 18:22 UT $\pm$ 9 minutes, with an apparent path
length of 2.06 $\pm$ 0.09 AU. This is again relatively late compared with
the flare start time (16:46 UT), while the associated CME was at the height 
of $\sim$\,16 R$_{\odot}$. An estimate based on the earliest observed onset
at 57.4 MeV provides the release time at $\sim$\,17:30 UT using a 1.2~AU path length, 
at which time the associated CME was at the height of $\sim$\,6~R$_{\odot}$. 
It is thus reasonable to assume that this SEP event was associated with the 
M5.7 flare at 16:46 UT and the CME first-observed at 17:30 UT. 
The derived electron release times of 17:11 UT (EPAM) and 17:18 UT
(EPHIN) support the earlier proton release time based on the first
observation of 57.4 MeV protons.

We note that in August 2005 the SOHO spacecraft was oriented so that
the ERNE view cones were pointing 45 degrees East of the Sun-Earth line, 
instead of the nominal 45 degrees West. This attitude may have prevented 
the observation of the first-arriving field-aligned particles, thus causing a 
delay in the onset times.

Again, no particle enhancements were observed in the vicinity of the third flare 
at 14:19~UT on 23 August, or during the ascent of the associated very fast halo CME. 
There was a drop of intensities at middle and high energies late on 23 August 
related to the ICME, and a small shock peak at lower energies at 
$\sim$\,07:30--09:30 UT on 24 August (Figure \ref{22aug2005}).
There was no indication of a new solar event visible in the electron
data at the time of the third flare-CME event, although there was a slight 
slow increase of near-relativistic electrons starting at 16--17 UT on 23 August.

During the time period of 22--23 August, 2005, the azimuth angle of the
IP magnetic field in GSE coordinates was predominantly in the range
[90$^{\circ}$, 270$^{\circ}$] (away from the Sun), although this period occurred in the
middle of a negative polarity sector. During the first and third flare-CME event, 
the field direction was clearly in this range. At the time of the second event, 
there were strong fluctuations in the field direction while the field strength 
remained fairly constant. Nevertheless, during the first and third flare-CME 
event the magnetic conditions near Earth were quite similar, and the missing
third SEP event cannot be explained by changes in local magnetic conditions.

On the other hand, at the time of the expected third SEP event, the two 
previous CMEs were still propagating between the Sun and Earth. Because the 
speed of the second CME was much higher than that of the first one (see 
Table \ref{table:3}), these CMEs were probably merged together (see also 
the discussion in Section \ref{sec:22aug}), and could have formed a strong obstacle 
for the propagation of particles associated with the third flare-CME event.

\subsection{Source Regions}

\subsubsection{24 November 2000}

The three X-class flares on 24 November 2000 were homologous, with a typical 
two-ribbon structure, and the plasmoid ejections or filament eruptions
and CMEs associated with each were observed to have almost the same direction 
\citep{takasaki2004}. Several studies on these events have suggested that the 
flares were triggered by newly emerging flux in the same part of the active region, 
which lead to reconnection and particle acceleration \citep{nitta2001,park2013}.
\citeauthor{takasaki2004} (\citeyear{takasaki2004}) also discovered that the rise times
of the flares got longer, the plasmoid velocities got smaller, and the separation
velocities of the flare ribbons became larger after each event. This may imply that
that the onsets of the flares occurred on larger flare loops in the later flares and that
the reconnection points were higher in the corona.  

A detailed study of the second flare-CME event also concluded that several plasmoids
were ejected during the flare but that they were merged as one CME, with the apparent 
velocity of the CME being larger than those of the plasmoids \citep{nishizuka2010}.
This suggested that the merged plasmoids were continuously accelerated as they 
were ejected into the IP space. 

%%%
\begin{figure}[!htbp]
   \centering
   \includegraphics[height=0.45\textwidth]{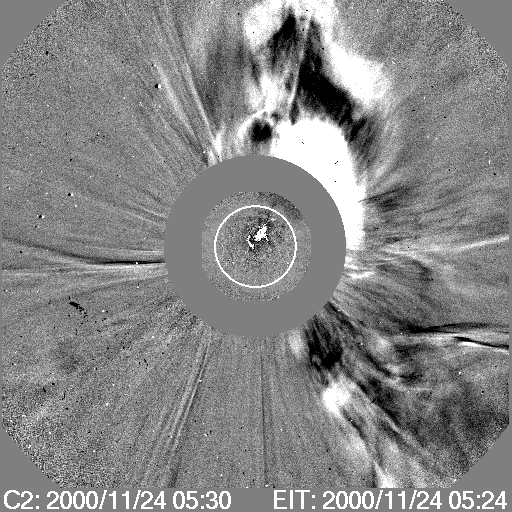}
   \includegraphics[height=0.45\textwidth]{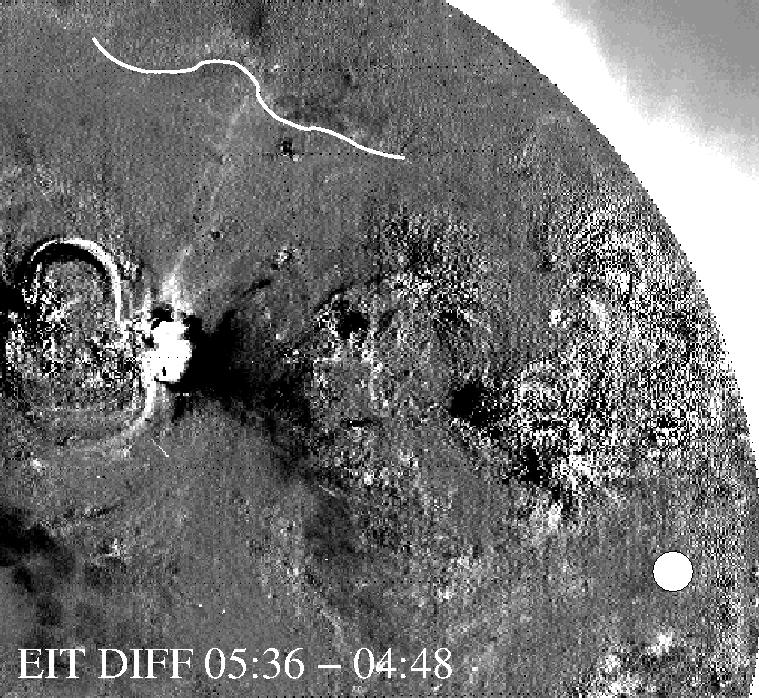}\\
   \includegraphics[height=0.45\textwidth]{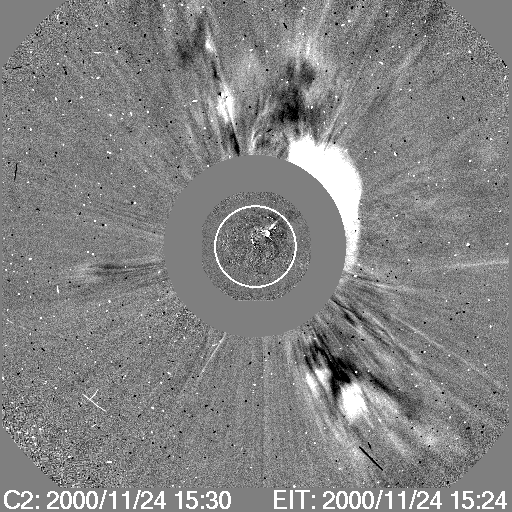}
   \includegraphics[height=0.45\textwidth]{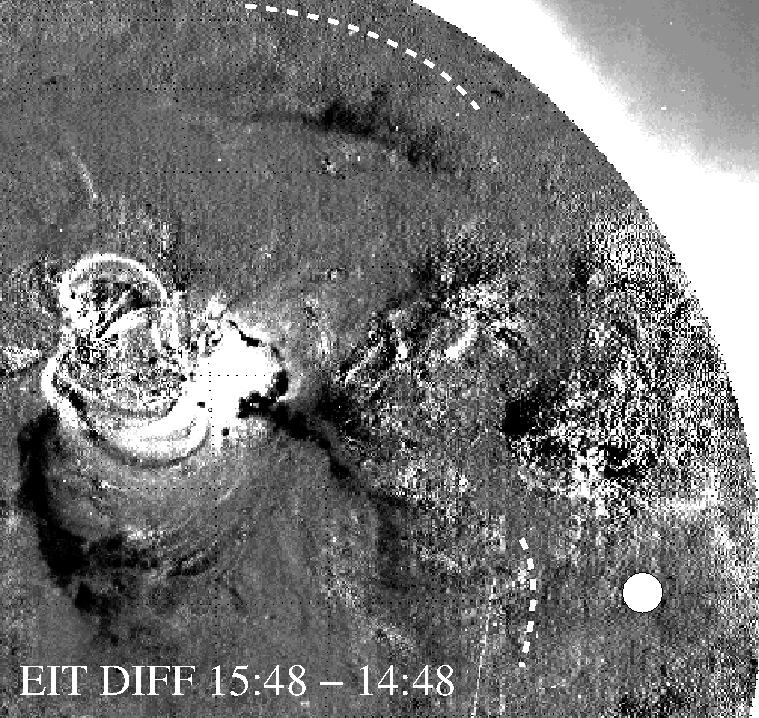}\\
   \includegraphics[height=0.45\textwidth]{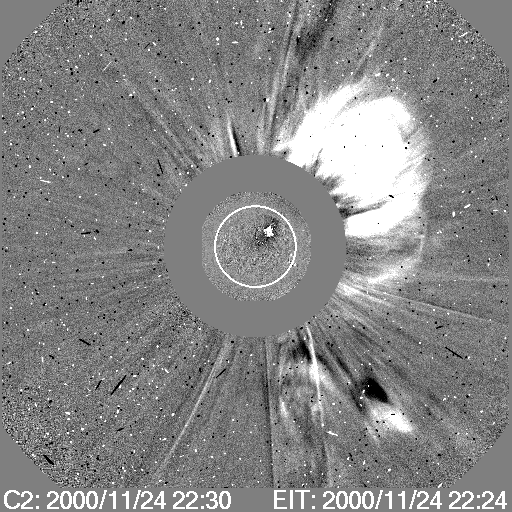}
   \includegraphics[height=0.45\textwidth]{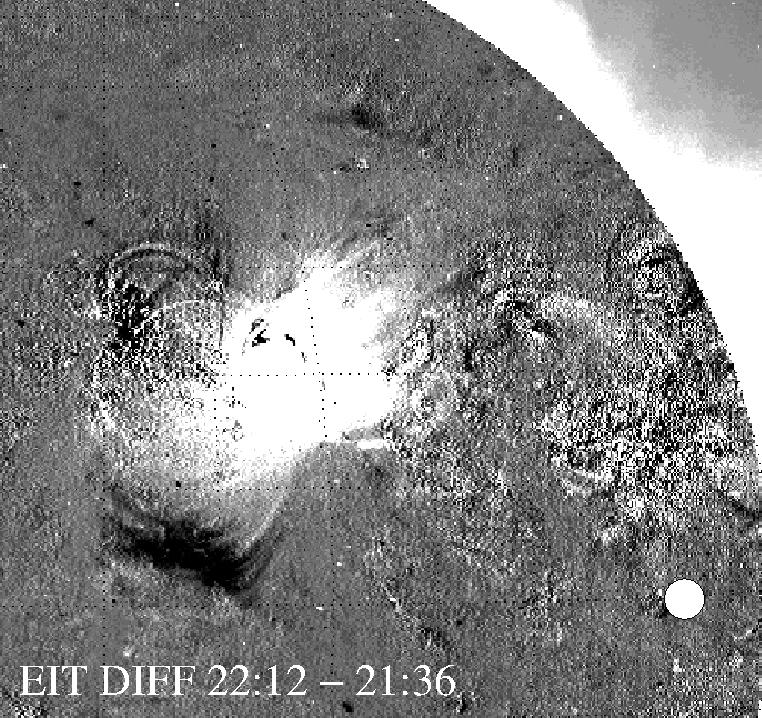}
   \caption{On the left: Difference images showing the coronal mass ejections 
  and the EUV dimming regions near the probable SEP injection times for the three 
  events on 24 November 2000 (SOHO/LASCO C2 and EIT images, from the CME catalog).
  On the right: EIT difference images (195 \AA) showing the locations of the 
  EUV waves (marked with white lines, the solid line indicates a clear front and the 
  dashed line more fuzzy structure). The white circle points to the location 
  of the nominal magnetic footpoint of the Earth. The thin black dashed-lines in the 
  EIT difference images are the solar grid plotted with 10 degree separation.}
 \label{CME24nov}%
   \end{figure}

We compared the large scale disturbances on the solar disk during each event on 24
November 2000 by analysing the SOHO/EIT difference images. All three events were 
associated with an EUV wave. Table \ref{table:4} gives the time of the EIT image
when the EUV wave was clearly visible. 
The partial-view EIT difference images in Figure \ref{CME24nov} show the EUV wave 
front locations with respect to the nominal magnetic footpoint of the Earth. The 
calculated footpoint longitudes for the three events were W57--W61, W50--W54, and 
W56--W60, respectively, with the given range indicating estimated uncertainty 
resulting from the uncertainty in the used average solar wind speed. The images 
show that only in the second event the wave looks to reach the magnetic footpoint, 
and in all events the wave propagation was probably affected by the active regions 
located in between the burst region and the magnetic footpoint. The third event 
does not show clear propagating wave fronts.

The only clear qualitative difference between the events was the extent of the 
dimming regions that appeared after the wave. In the first event two deep separate 
dimmed regions appeared. Also in the second event two regions appeared but they were 
spatially smaller. In the third event only one dimmed region appeared. 
The time separation from flare start to EUV wave detection was 17--21 minutes.

The associated CMEs looked similar at the calculated times of SEP injection 
(in the first two events when a SEP enhancement was observed) and at the estimated 
time of SEP injection for the third event (with no SEP enhancement), see 
Figure \ref{CME24nov}. Table \ref{table:4} gives also the estimated height of the 
CME at the calculated times of SEP injection.  In the later CME images there is
some indication that the third CME was less sharp-edged at the leading front 
and less spherical in intensity at larger distances.

\subsubsection{9--11 April 2001}

The three flares on 9--11 April 2001 had different X-ray intensities, from M2.3 to
X2.3 GOES class. The durations of the first and third flare were similar and they 
were much shorter than the second one. 

\begin{figure}[!htbp]
   \centering
   \includegraphics[height=0.45\textwidth]{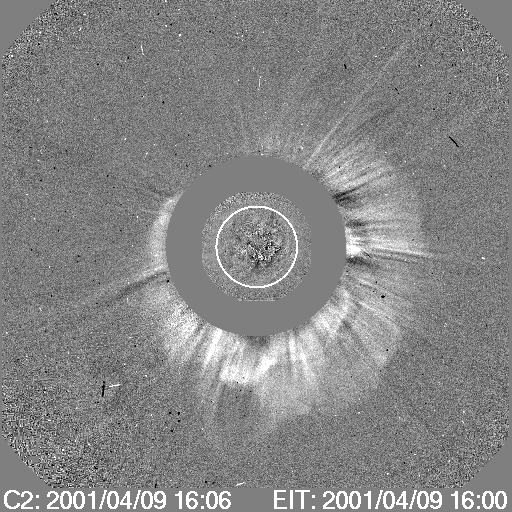}
   \includegraphics[height=0.45\textwidth]{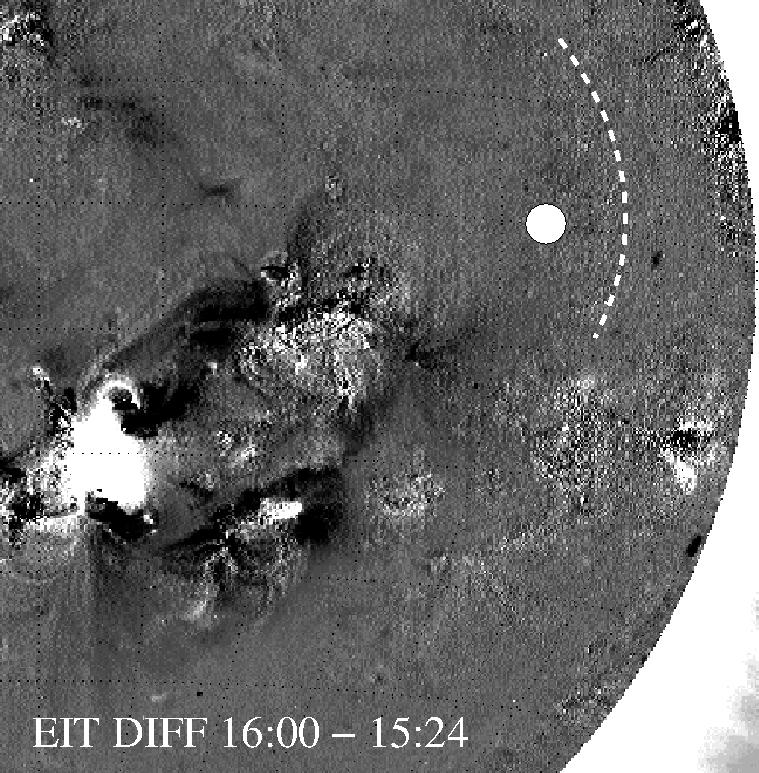}\\
   \includegraphics[height=0.45\textwidth]{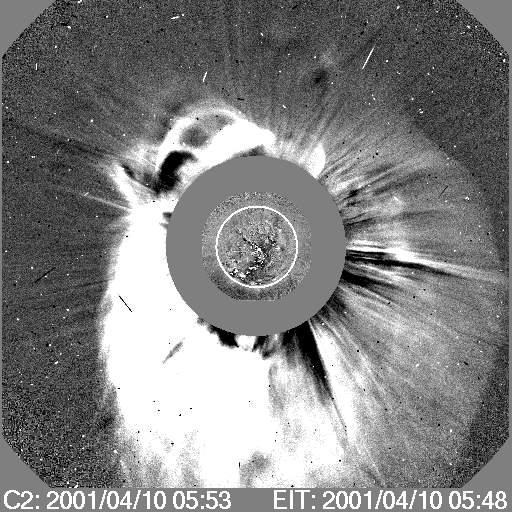}
   \includegraphics[height=0.45\textwidth]{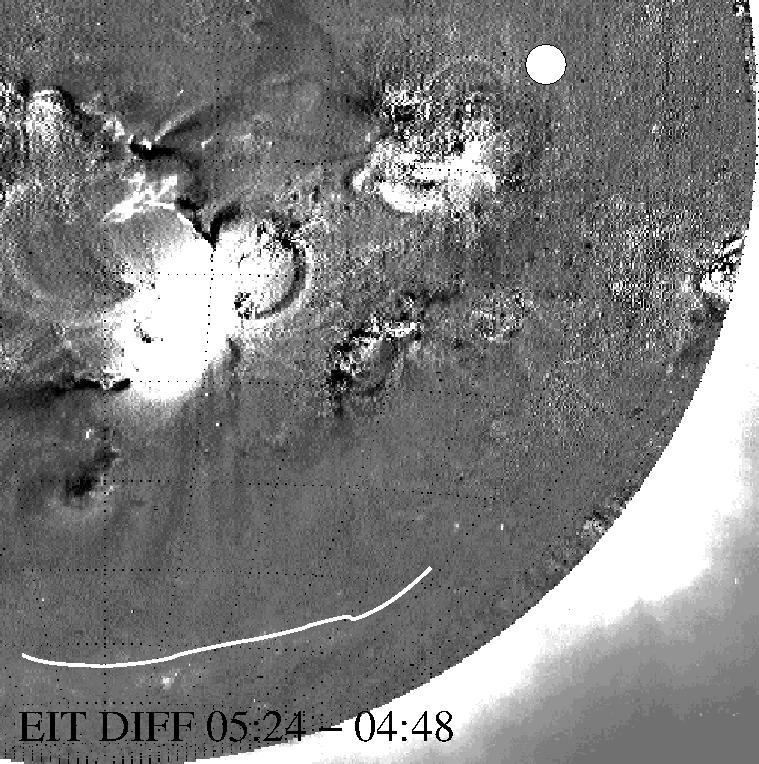}\\
   \includegraphics[height=0.45\textwidth]{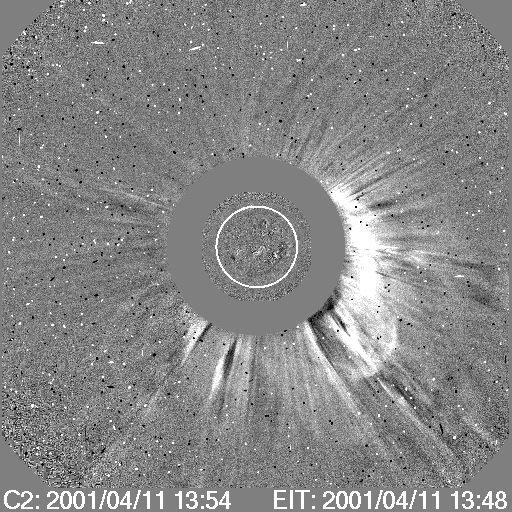}
   \includegraphics[height=0.45\textwidth]{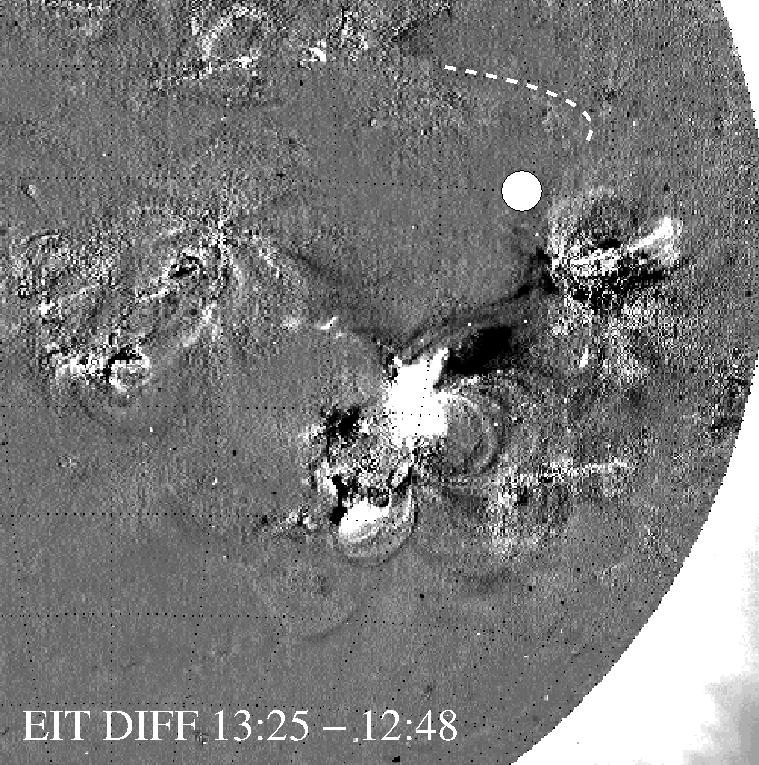}
   \caption{On the left: Difference images showing the coronal mass ejections 
  and the EUV dimming regions near the probable SEP injection times for the three events 
  on 9--11 April 2001 (SOHO/LASCO C2 and EIT images, from the CME catalog).
   On the right: EIT difference images (195 \AA) showing the locations of the 
  EUV waves (marked with white lines, the solid line indicates a clear front and the 
  dashed line more fuzzy structure). The white circle points to the location 
  of the nominal magnetic footpoint of the Earth. The thin black dashed-lines in the 
  EIT difference images are the solar grid plotted with 10 degree separation.
  } 
 \label{CME9apr}%
   \end{figure}

The events on 9--11 April 2001 were also associated with EUV waves 
(Table \ref{table:4}). The partial-view EIT difference images in Figure \ref{CME9apr} 
show the EUV wave front locations with respect to the nominal magnetic footpoint of 
the Earth. The calculated footpoint coordinates for the three events were W42--W46, 
W42--W46, and W34--W38, respectively. In the first and third events the wave looks to 
reach the magnetic footpoint, but in the second event the wave propagates most noticeably 
towards the South. In the second event another active region is also located in between 
the burst region and the magnetic footpoint, which may affect the wave propagation.

As in the 24 November 2000 case, the first two events showed deeper two-part dimmed 
regions and the dimming in the third was less intense and spatially smaller. The 
time of the EUV wave listed for the second event in the row (10 April, at 05:24 UT) is 
later than what is reported in \citeauthor{miteva2014} (\citeyear{miteva2014}). This is 
because we list the main wave-like brightening and not the preceding faint transients 
observed near the erupting active region. The time separation from the flare start to 
the EUV wave detection was 16--18 minutes.
 
The appearances of the associated CMEs were again similar. Figure \ref{CME9apr} 
shows the LASCO C2/EIT difference images at the time of the calculated SEP injection 
time (first event), and 0.8 hours after the flare start (second event), and 1.0 hour 
after the flare start (third event), to show the later two at the same phase of 
propagation as the first one. The SEP analysis, however, suggests that the SEP 
injection occurred at a much later time in the second event, about 1.7 hours after 
the flare start (Table \ref{table:4}).  

\begin{figure}[!htbp]
   \centering
   \includegraphics[height=0.45\textwidth]{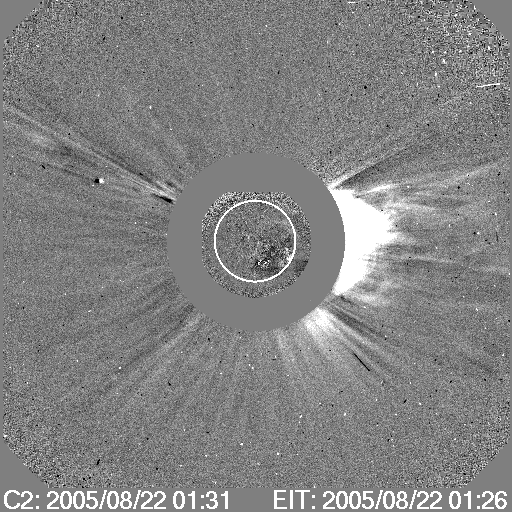}
   \includegraphics[height=0.45\textwidth]{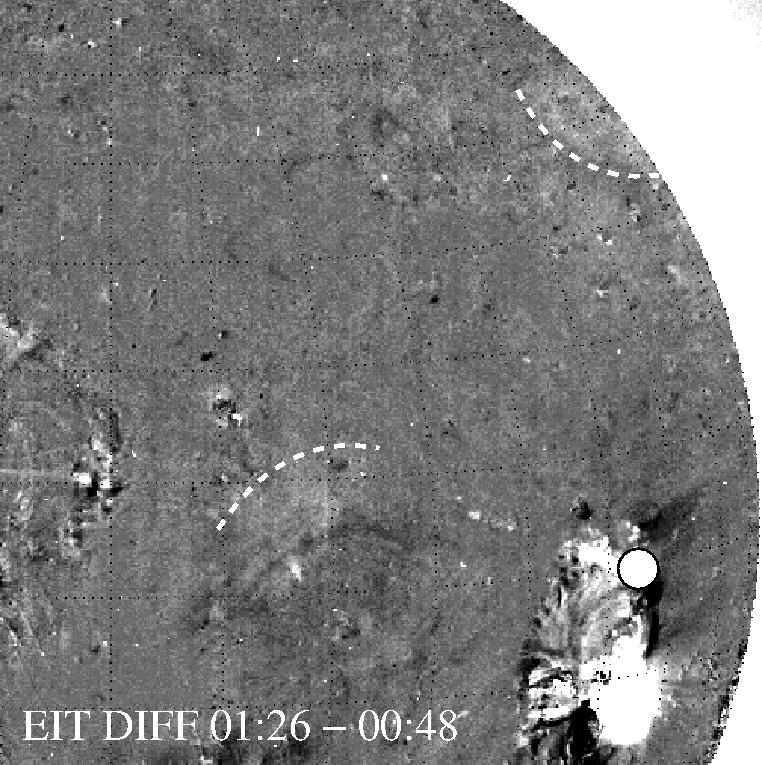}\\
   \includegraphics[height=0.45\textwidth]{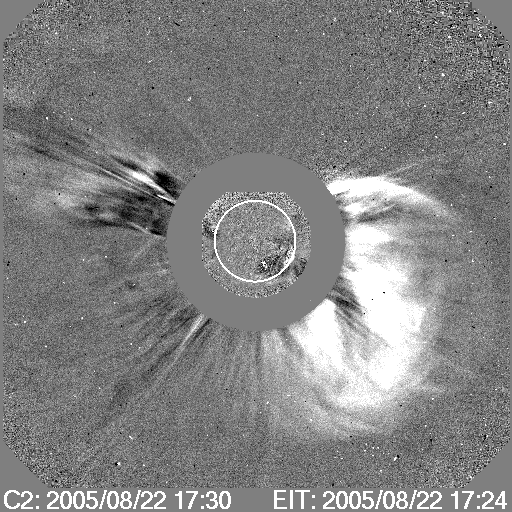}
   \includegraphics[height=0.45\textwidth]{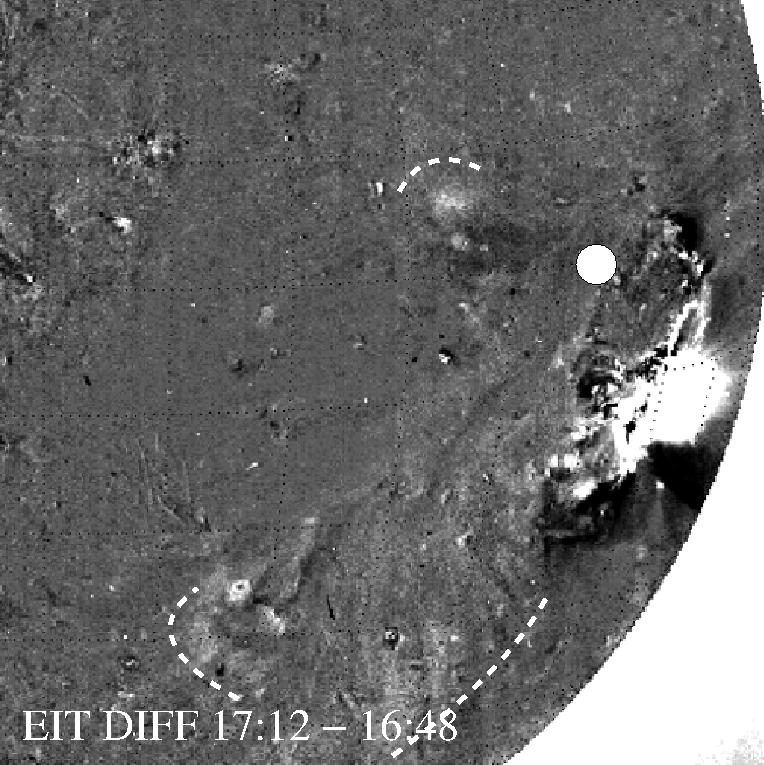}\\
   \includegraphics[height=0.45\textwidth]{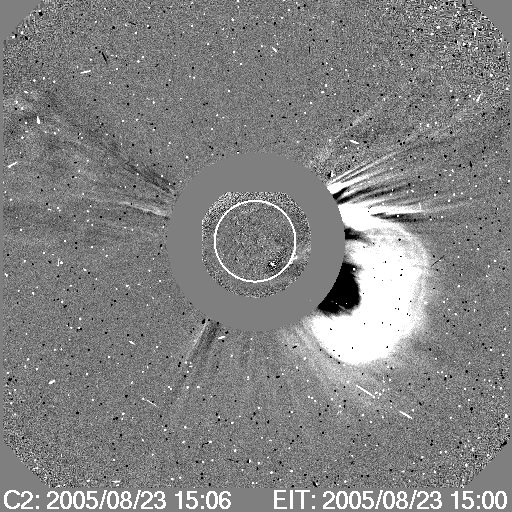}
   \includegraphics[height=0.45\textwidth]{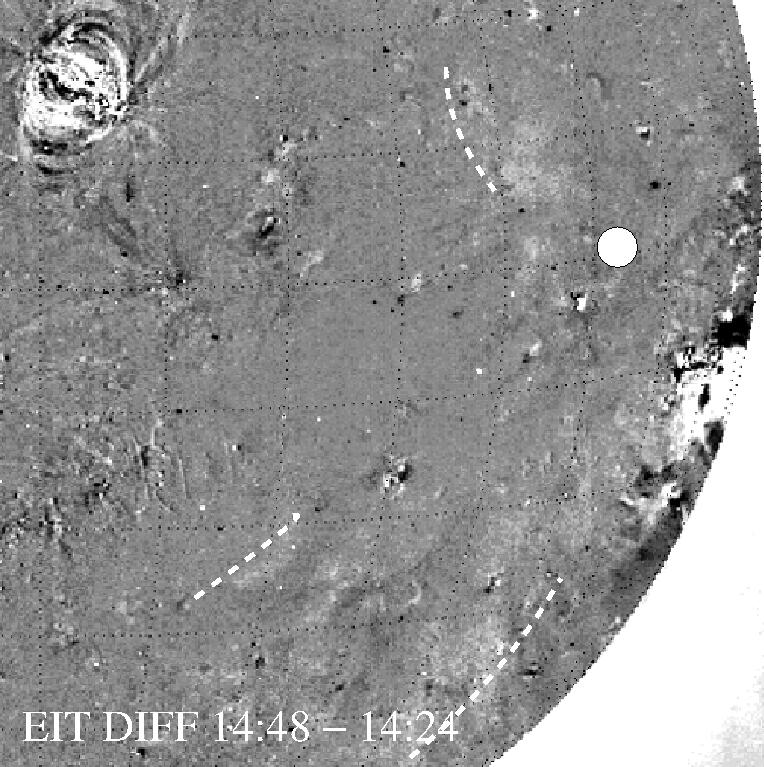}
  \caption{On the left: Difference images showing the coronal mass ejections and 
  the EUV dimming regions near the probable SEP injection times for the three events on 
  22--23 August 2005 (SOHO/LASCO C2 and EIT images, from the CME catalog).
  On the right: EIT difference images (195 \AA) showing the locations of the 
  EUV waves (marked with white lines, the solid line indicates a clear front and the 
  dashed line more fuzzy structure). The white circle points to the location 
  of the nominal magnetic footpoint of the Earth. The thin black dashed-lines in the 
  EIT difference images are the solar grid plotted with 10 degree separation.
  } 
 \label{CME22aug}%
   \end{figure}

\subsubsection{22--23 August 2005}

The active region that produced the observed flares on 22--23 August 2005 emerged
in the middle of a small coronal hole and formed an anemone-like structure \citep{asai09}.
The region showed a reversed polarity orientation (in violation to the Hale-Nicholson 
magnetic polarity law) which made it statistically more likely to erupt. The magnetic
flux and soft X-ray intensity were observed to grow gradually (see Figure 3 in 
\citeauthor{asai09}, \citeyear{asai09}), and the three M-class flares were produced 
before the active region rotated behind the West limb. 

Each event was associated with an EUV wave (Table \ref{table:4}). 
The partial-view EIT difference images in Figure \ref{CME22aug} show the EUV wave 
front locations with respect to the nominal magnetic footpoint of the Earth. 
The calculated footpoint coordinates for the three events were W52--W55, 
W47--W51, and W53--W57, respectively. In the first event the footpoint was basically 
inside the flaring active region and in the second and third events parts of the EUV 
wave look to cross the magnetic footpoint region.
No obvious qualitative differences were found between the events, although the 
waves were in general less intense than in the 24 November 2000 and 9--10 April 2001 
events. The time separation from flare start to the EUV wave detection was 26--29 minutes,
which is more than in the two other cases.

Also the associated CMEs were similar. Figure \ref{CME22aug} shows the 
LASCO C2/EIT difference images near the time of the calculated SEP injection time 
(first event), and 0.7 hours after the flare start (second event), and 0.8 hours 
after the flare start (third event), to show the later two at the same phase of 
propagation as the first one. The SEP analysis again suggests that the SEP injection 
occurred at a much later time in the second event, about 1.6 hours after the 
flare start (Table \ref{table:4}).

\section{Results}

We have analysed triple solar events, where intense flares occur in the 
same active region one after the other (three in a row), each producing an 
EUV wave, a fast halo-type CME, and a propagating shock wave visible as an 
interplanetary radio type II burst, and with energetic particles associated 
with at least the first two events.
Based on the preliminary event list presented in Table \ref{table5} (Appendix), 
and after using the additional selection criteria described in Section \ref{sec:data}, 
three sequences of events were selected for detailed analyses. These events were 
observed in the years 2000, 2001, and 2005, and they showed the special feature 
that while the first two solar bursts were associated with a SEP enhancement, 
as required, the third was not.

The velocity dispersion analysis (VDA) for the first burst in each sequence 
gave a solar proton injection time of 0.4--0.7 hours after the flare start, 
at which time the CME was at a height of about 2.5--4.0 R$_{\odot}$. 
For the second bursts in the row, a delay in the SEP injection time was found 
in two cases (10 April 2001 and the 22 August 2005 events). 
A delay in the proton onset times at 1 AU was also reported by 
\citeauthor{miteva2014} (\citeyear{miteva2014}). 
This means that the energetic particles could have been injected some 1.6--1.7 hours 
after the flare, when the CMEs were at heights of about 16.0--19.0 R$_{\odot}$. 
We note that no delay was observed in the 24 November 2000 case. There the 
second SEP injection occurred 0.6 hours after the flare start, at which 
time the CME was at a height of about 3.0 R$_{\odot}$.      

Based on the simple assumption of a 1.2 AU path length, the electron injection times 
of the first event of the sequences in 2000 and 2001 were consistent with the proton 
observations. In the sequence of 2005, the agreement was still reasonable, although the 
derived electron injection times were somewhat earlier than the proton injection time. 
The agreement was good also in the second event in 2000, while in the two other cases 
the electron injection seemed to be earlier than proton injection, but still relatively 
late compared to the flare.

All the nine flare-CME events, in the sequences of three triple events, were 
associated with large-scale disturbances on the solar disk, observed in EUV. 
The exact start times of the EIT (EUV) waves could not be determined due to the 
low EIT image cadence, but the listed wave observation times (Table \ref{table:4}) 
show similar time scales in all events. The EUV waves were all followed by dimmings, 
which typically indicate the depletion of material. We observed that the dimmings in the 
later events (after the first event in the row) were spatially smaller and less deep, 
which could mean less ejected material within the CMEs. 

All the events were associated with radio type III bursts. However, in each sequence 
only the first two bursts continued down to the plasma frequency level near Earth, 
to about 30 kHz. In all three sequences the third type III bursts ended at a higher 
frequency, near 200 kHz (24 November 2000), 60 kHz (11 April 2001), and 50--80 kHz 
(23 August 2005). These type III cut-off periods lasted for some time, about one day, 
by which time the associated earlier-launched shocks had arrived to Earth.  

Only in April 2001, the inspected {\it in-situ} magnetic field data indicated a 
possibility that the magnetic field conditions near Earth would have been such that 
propagation of particles associated with the third flare-CME event could have been obstructed. 
These {\it in-situ} observations, however, do not preclude the possibility that the two 
previous CME-shock fronts could not have blocked the path of the particles associated 
with the third flare-CME event also for the two other sequences.

The first two propagating shocks that started on 24 November 2000 were recorded to 
arrive near Earth on 26 November. The arrival of the third shock is uncertain, but 
based on earlier studies it is possible that the observed shocks (and CMEs) had merged
during propagation. The three propagating shocks that started on 9, 10, and 11 April 2001 
had corresponding shock arrivals near Earth. The first two CMEs/shocks probably 
merged in some extent as we note that several shocks were identified at the 
estimated arrival times of the first two. The third CME on 11 April was identified 
by IPS measurements, which verifies the height-time evolution. The first two propagating 
shocks that started on 22 August 2005 were recorded as arriving shocks near Earth 
on 24 August. Again, merging of the CMEs and shocks looks probable. 
The arrival near Earth of the third shock, which started on 23 August, was 
uncertain but a disturbance was reported to arrive on 25 August, which fits with 
the estimated height-time evolution.

\section{Discussion and Conclusions}

We have analysed three sequences of triple flare and CME events on 24--26 November 2000, 
9--13 April 2001, and 22--25 August 2005, and their solar energetic particle (SEP) 
associations. While the propagating shocks in each event could be traced by their 
interplanetary type II radio emissions, only the first two events in each sequence 
were associated with SEP enhancements near Earth. The observed radio emission frequencies 
depend only on the local plasma densities, and the standard atmospheric density models 
assume spherically symmetric and uniform density structures. 
However, the actual density structures in the interplanetary medium are both complex 
and variable \citep{reiner2001}. For example, the plasma density corresponding to 
the frequency of 60 kHz can occur at distances of 45--215 R$_{\odot}$, depending on 
the ecliptic longitude (see, for example Figure 9 in \citeauthor{lecacheux1989}, 
\citeyear{lecacheux1989}). Therefore the obtained type II burst heights can sometimes 
be questioned, especially at large distances from the Sun. However, shock arrivals
near Earth can be registered and compared with the earlier-observed height-time 
evolution of the propagating shocks. In most of our events shock arrivals were 
observed.  

Since all of the events showed radio type III emission ({\it i.e.}, fast-drift bursts 
caused by accelerated electron beams), we had a closer look on their features. 
Our observation that the type III bursts associated with the third 
flare-CME events did not reach the near-Earth plasma frequencies deserves attention.
It is known that during the propagation of type III radio bursts, the radiation 
suffers refraction in density gradients and scattering by inhomogeneities \citep{dulk2000}. 
Recent numerical kinetic simulations by \citeauthor{reid2015} (\citeyear{reid2015}) 
have shown that the stopping frequency of type III bursts can be affected by 
several conditions: the expansion of the guiding magnetic flux tubes, the density 
and the spectral index of the injected electron beams, as well as the large-scale 
density fluctuations in the background electrons. It is likely 
that our type III cut-off periods were caused because the Langmuir wave production 
was either stopped or the beam paths were diverted before they reached the Earth. 
The higher cut-off frequencies could also be explained by higher plasma densities 
near Earth.
However, based on ACE/Solar Wind Electron, Proton, and Alpha Monitor 
(SWEPAM: \citeauthor{McComas98}, \citeyear{McComas98}) solar wind data, 
there is no indication of increased plasma density during the 24 November 2000 and 
23 August 2005 events. In this respect the event of 11 April 2001 is
different. Due to the shock arriving at Earth simultaneously with the
occurrence of the third flare-CME event of this sequence, there is a
sudden increase in the solar wind proton density. 

The results from our analyses of the three triple-event sequences suggest that electron 
beams cannot propagate directly through plasma structures created by at least two earlier 
CMEs and their associated propagating shocks. The same could be expected for energetic 
protons. An alternative explanation for not observing SEP enhancement associated with 
the third flare-CME event would be that the earlier CMEs have wiped out most of 
the seed particles, with very little particles left for the 
third CME to accelerate. 
This possibility is supported by the fact that in the sequences of
events in November 2000 and August 2005, no flares were observed
between the second and third flare-CME events (from the same active
region as the flares of the sequence) and only one C-class flare was
reported between the second and third events in the case of April 2001
sequence. Thus, the seed particle population may have been significantly
depleted before the third events of the sequences. In April 2001 there
was also a fourth event (without IP type II burst), which was
associated also with a strong SEP event. However, in this case the
event was preceded by three flares (C7.7, B7.2, and M1.3) from the
same active region, which could have filled in the seed particle
reservoir.

By comparing the flare, EUV wave, CME, and propagating shock wave characteristics we 
could not find any reason why SEPs would not have been accelerated in all of these events. 
We suggest that the earlier-launched CMEs and the CME-driven shocks either reduced 
the seed particle population and thus led to inefficient particle acceleration, or 
that the propagation paths of both electron beams and SEPs were changed 
by the effect of the previous eruptions, so that they could not reach the 
Earth-connected field lines, or that at some point of their propagation 
between the Sun and Earth they encountered disturbed magnetic fields preventing 
their propagation to Earth. For example, during the first two events in each sequence 
the geometry of the shock could have been such that it intercepted the interplanetary 
magnetic field lines connecting to the Earth, while during the third event particles 
were injected in magnetic field lines not reaching the Earth. Hence,  
particles were not detected near Earth even if the shocks themselves were 
observed to arrive. 

%%%%

\begin{acks}
We are grateful to all the individuals who have contributed in creating and updating 
the various solar event catalogues. The CME catalog is generated and maintained at 
the CDAW Data Center by NASA and the Catholic University of America in cooperation 
with the Naval Research Laboratory. The {\it Wind} WAVES radio type II burst catalog 
has been prepared by Michael L. Kaiser and is maintained at the Goddard Space Flight Center. 
We acknowledge the use of SEPServer as the source of the electron data.
Ian Richardson and Hilary Cane are thanked for the ICME and SEP catalogs. 
SOHO is a project of international cooperation between ESA and NASA. 
\end{acks}

%%%%%%%%%%%%%%%%%%%%%%%%%%%%%%%%%%%%

\appendix 

The consecutive multiple solar eruptions found from 1997--2012 are listed in 
Table \ref{table5}.  

\begin{table}[!t]
\caption{Close-by events that show a flare, a CME, and a decametric-hectometric 
(DH) type II burst, found from the list of {\it Wind}/WAVES type II bursts and CMEs at 
 \url{http://cdaw.gsfc.nasa.gov/CME_list/radio/waves_type2.html}. 
The existence of a listed SEP event has been added to each entry (yes).
These listed SEP events are from \citeauthor{cane2010} (\citeyear{cane2010})
and \citeauthor{richardson2014} (\citeyear{richardson2014}).  
The information on the SEP intensities in the non-listed cases are given as 
rise (flux increasing but not listed as an event), fall (flux decreasing), 
r/f and f/r (both rise and fall observed, in the given order), flat (constant and 
unchanging, can be background level or on top of a long-duration event). 
The observed type III burst cut-offs are also listed in the entries (yes). 
In this column 2f$_p$ indicates that the bursts end near the second harmonic of 
the local plasma frequency, AKR stands for auroral kilometric radiation that 
masks any cut-offs if they exist, and `--' stands for no observed cut-off.
The dates with an asterisk indicate the events that are analysed in this paper. 
}             
\label{table5}               
\begin{tabular}{l l r r r r r l l}
\hline       
Date        & DH      & Start     & NOAA  & GOES   & CME   & CME    & SEP  & Type \\
            & type II & freq.     & AR    & flare  & width & speed  & event& III \\
            & start   &           &       & class  &       & fitted &      & cut-off\\
yy/mm/dd  &(UT)     & (kHz)     &       &        & (deg) & (km/s) &      & \\
\hline
1997/11/03  & 05:15  & 14000  & 8100  & C8.6  & 109  & 227    & flat & 2f$_p$\\
1997/11/03  & 10:30  & 14000  & 8100  & M4.2  & 122  & 352    & yes  & 2f$_p$\\
1997/11/04  & 06:00  & 14000  & 8100  & X2.1  & 360  & 785    & yes  & -- \\
1997/11/06  & 12:20  & 14000  & 8100  & X9.4  & 360  & 1556   & yes  & -- \\
\hline
2000/11/24* & 05:10  & 14000  & 9236  & X2.0  & 360  & 1289  & yes  & -- \\
2000/11/24* & 15:25  & 14000  & 9236  & X2.3  & 360  & 1245  & yes  & -- \\
2000/11/24* & 22:24  &  4000  & 9236  & X1.8  & 360  & 1005  & fall & yes\\
2000/11/25  & 19:00  &  6000  & 9236  & X1.9  & 360  & 671   & rise & yes\\
2000/11/26  & 17:00  & 14000  & 9236  & X4.0  & 360  & 980   & rise & yes\\
\hline
2001/03/27  & 02:35  & 6000   & 9393  & C7.3  & 60   & 300   & rise & -- \\
2001/03/27  & 15:00  & 4000   & 9393  & C5.6  & 66   & 340   & rise & -- \\
2001/03/29  & 10:12  & 4000   & 9393  & X1.7  & 360  & 942   & yes  & -- \\
\hline
2001/04/09* & 15:53  & 12000  & 9415  & M7.9  & 360  & 1192  & yes  & -- \\
2001/04/10* & 05:24  & 14000  & 9415  & X2.3  & 360  & 2411  & yes  & -- \\
2001/04/11* & 13:15  & 14000  & 9415  & M2.3  & 360  & 1103  & fall & yes\\
2001/04/12  & 10:20  & 14000  & 9415  & X2.0  & 360  & 1184  & yes  & -- \\
\hline
2002/07/15  & 21:15  & 5000   & 10030 & M1.8  &$>$188& 1300  & flat & -- \\
2002/07/17  & 07:30  & 3500   & 10030 & M8.5  & 177  & 716   & r/f & --\\
2002/07/18  & 07:55  & 14000  & 10030 & X1.8  & 360  & 1099  & fall & -- \\
\hline
2002/11/09  & 13:20  & 14000  & 10180 & M4.6  & 360  & 1838  & yes & -- \\
2002/11/10  & 03:20  & 3000   & 10180 & M2.4  & 282  & 1670  & r/f & --\\
2002/11/11  & 16:15  & 14000  & 10180 & M1.8  & 93   & 1083  & fall & 2f$_p$\\
\hline
2003/05/27  & 23:12  & 14000  & 10365 & X1.3  & 360  & 964   & yes & -- \\
2003/05/28  & 01:00  & 1000   & 10365 & X3.6  & 360  & 1366  & yes & -- \\
2003/05/29  & 01:10  & 13000  & 10365 & X1.2  & 360  & 1237  & flat& -- \\
\hline
2004/01/06  & 00:00  & 14000  & 10537 & M5.8  & 166  & 1469  & flat & AKR\\  
2004/01/07  & 04:15  & 14000  & 10537 & M4.5  & 171  & 1581  & flat & AKR\\
2004/01/07  & 10:35  & 14000  & 10537 & M8.3  & 182  & 1822  & yes  & AKR\\
\hline
\end{tabular}
\end{table}

\begin{table}[!t]                     
\begin{tabular}{l l r r r r r l l}
\hline       
Date        & DH      & Start     & NOAA  & GOES   & CME   & CME    & SEP  & Type \\
            & type II & freq.     & AR    & flare  & width & speed  & event& III \\
            & start   &           &       & class  &       & fitted &      & cut-off\\
yy/mm/dd    &(UT)     & (kHz)     &       &        & (deg) & (km/s) &      & \\
\hline
2004/04/06  & 13:05  & 8000   & 10588 & M2.4  & 360  & 1368  & rise & 2f$_p$\\
2004/04/08  & 10:25  & 3000   & 10588 & C7.4  & 360  & 1068  & rise & -- \\
2004/04/08  & 13:30  & 6000   & 10588 & C1.3  & 92   & 959   & rise & -- \\
\hline
2004/11/07  & 16:25  & 14000  & 10696 & X2.0  & 360  & 1759  & yes & -- \\
2004/11/09  & 17:35  & 14000  & 10696 & M8.9  & 360  & 2000  & yes & -- \\
2004/11/10  & 02:25  & 14000  & 10696 & X2.5  & 360  & 3387  & yes & -- \\
\hline  
2004/12/29  & 16:35  & 14000  & 10715 & M2.3  & 140  & 774   & flat & 2f$_p$\\
2004/12/30  & 23:45  & 5000   & 10715 & M4.2  & 360  & 1035  & flat & 2f$_p$\\
2005/01/01  & 00:45  & 14000  & 10715 & X1.7  & 360  & 832   & flat & -- \\
\hline                  
2005/01/15  & 06:15  & 14000  & 10720 & M8.6  & 360  & 2049  & yes  & -- \\
2005/01/15  & 23:00  & 3000   & 10720 & X2.6  & 360  & 2861  & yes  & -- \\
2005/01/17  & 09:25  & 14000  & 10720 & X2.0  & 360  & 2094  & yes  & -- \\
2005/01/17  & 09:43  & ?      & 10720 & X3.8  & 360  & 2547  & yes  & -- \\
2005/01/19  & 09:20  & 5300   & 10720 & X1.3  & 360  & 2020  & fall & -- \\
2005/01/20  & 07:15  & 14000  & 10720 & X7.1  & 360  & 882   & yes  & -- \\
\hline
2005/08/22* & 01:30  & 8000   & 10798 & M2.6  & 360  & 1194  & yes  & -- \\
2005/08/22* & 17:15  & 12000  & 10798 & M5.6  & 360  & 2378  & yes  & -- \\
2005/08/23* & 15:00  & 13000  & 10798 & M2.7  & 360  & 1929  & fall & yes \\
\hline
2005/09/09  & 19:45  & 10000  & 10808 & X6.2  & 360  & 2257  & fall & -- \\
2005/09/10  & 21:45  & 14000  & 10808 & X2.1  & 360  & 1893  & flat & 2f$_p$\\
2005/09/11  & 13:10  & 3000   & 10808 & M3.0  & 360  & 1922  & f/r & AKR\\
2005/09/13  & 20:20  & 1100   & 10808 & X1.5  & 360  & 1866  & yes  & -- \\
\hline
2011/09/22  & 11:05  & 14000  & 11302 & X1.4  & 360  & 1905  & yes & -- \\
2011/09/24  & 12:50  & 16000  & 11302 & M7.1  & 360  & 1915  & flat& 2f$_p$ \\
2011/09/25  & 05:30  & 16000  & 11302 & M7.4  & 193  & 788   & rise& yes \\
\hline
2012/03/07  & 01:00  & 16000  & 11429 & X5.4  & 360  & 2684  & yes & -- \\
2012/03/09  & 04:10  & 14000  & 11429 & M6.3  & 360  & 950   & r/f & yes\\
2012/03/10  & 17:55  & 14000  & 11429 & M8.4  & 360  & 1296  & flat & 2f$_p$\\
\hline
2012/07/05  & 22:40  & 3000   & 11515 & M1.6  & 94   & 980   & yes & -- \\
2012/07/06  & 23:10  & 16000  & 11515 & X1.1  & 360  & 1828  & yes & -- \\
2012/07/08  & 16:35  & 16000  & 11515 & M6.9  & 157  & 1495  & yes & -- \\
\hline                  
\end{tabular}
\end{table}

%%%%%%%%%%%%%%%%%%%%%%%%%%%%%%%%%%%%

% Without BibTeX 

%%%%%%%%%%%%%%%%%%%%%%%%%%%%%%%%%%%%%%%%%%%%%%%%%%%%%%%%%%%%%%%%%%%%%%%%%%%%%%%%%%%%%%%%%%%%%%%%%

\end{article}
\end{document}